# CMISR: Circular Medical Image Super-Resolution

Honggui Li[1*], Nahid Md Lokman Hossain[2], Maria Trocan[3], Dimitri Galayko[4], Mohamad Sawan[5,6]
[1,2]School of Information Engineering, Yangzhou University, Yangzhou 225127, China
[3]LISITE Research Lab, Institut Supérieur d'Électronique de Paris, Paris 75006, France
[4]Laboratoire d'Informatique de Paris 6, Sorbonne University, Paris 75020, France
[5]Polystim Neurotech Laboratory, Polytechnique Montreal, Montreal H3T1J4, Canada
[6]School of Engineering, Westlake University, Hangzhou 310024, China
[1]hgli@yzu.edu.cn, [2]mh23083@stu.yzu.edu.cn, [3]maria.trocan@isep.fr,
[4]dimitri.galayko@sorbonne-universite.fr, [5]mohamad.sawan@polymtl.ca, [6]sawan@westlake.edu.cn
[*]Corresponding Author

**Abstract**: Classical methods of medical image super-resolution (MISR) utilize open-loop architecture with implicit under-resolution (UR) unit and explicit super-resolution (SR) unit. The UR unit can always be given, assumed, or estimated, while the SR unit is elaborately designed according to various SR algorithms. The closed-loop feedback mechanism is widely employed in current MISR approaches and can efficiently improve their performance. The feedback mechanism may be divided into two categories: local feedback and global feedback. Therefore, this paper proposes a global feedback-based closed-cycle framework, circular MISR (CMISR), with unambiguous UR and advanced SR elements. Mathematical model and closed-loop equation of CMISR are built. Mathematical proof with Taylor-series approximation indicates that CMISR has zero recovery error in steady-state. In addition, CMISR holds plug-and-play characteristic that fuses model-based and learning-based approaches and can be established on any existing MISR algorithms. Five CMISR algorithms are respectively proposed based on the state-of-the-art open-loop MISR algorithms. Experimental results with three scale factors and on three open medical image datasets show that CMISR is superior to MISR in reconstruction performance and is particularly suited to medical images with strong edges or intense contrast.

## 1 Introduction

Medical imaging is one of the important technique foundations for medical disease diagnosis [1-2]. Classical medical imaging methods include magnetic resonance imaging (MRI), computed tomography (CT), X-ray radiography, ultrasound imaging, optical imaging, hyperspectral imaging, endoscopic imaging, and so on [3-7]. On the one hand, medical imaging possesses the following advantages: low radiation, low injury, low latency, low cost, non-invasion, etc. On the other hand, medical imaging possesses the following disadvantages: low-quality (LQ), low-resolution (LR), low-definition (LD), low contrast, gray value, etc. Hence, it is necessary for researchers to study medical image super-resolution (MISR) reconstruction before further processing, analysis, classification, recognition, and understanding [8-10].

MISR is a special branch of general image super-resolution (ISR) that mainly focuses on the super-resolution (SR) of natural images [11-18]. General ISR may be categorized into three classes: classical ISR, blind ISR, and real-world ISR. All algorithms of general ISR can be modified for MISR by considering the LQ characteristic of medical images. Some hereafter-mentioned state-of-the-art general ISR algorithms are enumerated as follows: hybrid attention transformer (HAT) combines both channel attention and window-based self-attention schemes [14]; SR transformer (SRFormer) takes advantage of permuted self-attention to seek an appropriate balance between channel and spatial information [15]; dual aggregation transformer (DAT) fuses features across spatial and channel dimensions [16]; efficiently mixed transformer (EMT) wields pixel mixer to replace self-attention and enhance local knowledge aggregation [17]; image restoration using shifted-window transformer (SwinIR) consists of three parts: shallow feature extraction, deep feature extraction and high-quality (HQ) image reconstruction [18].

Traditional MISR approaches may be divided into two categories: model-based approach and learning-based approach [19-23]. The former employs a known mathematical model, such as image prior, total variance, or sparse representation, is suitable for any image datasets, and holds a huge computing burden of iterative resolving. The latter doesn't employ an existing mathematical model but learns one, is suitable for a particular image dataset, and holds fast inference speed after longtime training with high computing complexity. The current development direction of MISR is to amalgamate the two approaches. For instance, a denoising prior model is obtained by a learning-based approach and is combined with a model-based approach in the form of plug-and-play [23].

Conventional MISR approaches make use of a serial open-loop architecture with an implicit under-resolution (UR) unit and an explicit SR unit [8-23]. The UR unit degrades HQ images to LQ images while the SR unit recovers HQ images from LQ images. UR unit can always be given, assumed, or estimated and SR is minutely designed depending on different MISR algorithms. For classical MISR, UR is a known down-sampling (DS) procedure, such as nearest, bilinear, trilinear, cubic, bicubic, or area interpolation. For blind MISR or real-world MISR, UR consists of unknown down-sampling and degradation subunits, and it can be guessed or calculated. Image DS and degradation processes are essential prior knowledge that provides the potential of improving MISR performance.

Classical MISR approaches utilize an open-loop feedforward mechanism to upgrade performance [24]. For instance, a residual network (ResNet) uses a skip link from the input to the output of the network unit [24]. At the same time, the closed-loop feedback (FB) mechanism is widely adopted in MISR [8-23]. FB mechanism efficiently lifts MISR performance via transferring an error term from the front to the back. For example, global negative FB is used in the iterative optimization of model-based MISR; global negative FB is used in the training phase of deep learning to update weights of deep neural networks; local FB is used in deep recurrent neural networks to achieve long-term and short-term memory performance. In order to enhance the performance of the existing MISR methods, it is necessary for scholars to change classical open-loop MISR architecture to closed-loop architecture.

Therefore, this paper proposes a circular closed-loop MISR (CMISR) framework with an explicit UR element and a perfect SR element in the form of plug-and-play which blends model-based and learning-based approaches.

The key innovations of this paper are listed as follows:
(1) closed-loop CMISR infrastructure with explicit UR and SR units compared with classical open-loop MISR with implicit UR and explicit SR;
(2) plug-and-play attribution which fuses model-based and learning-based methods and can be established on any existing MISR algorithms;
(3) detailed mathematical proof of zero recovery error with Taylor series expansion and high-order term discarding;
(4) significant performance improvement in peak signal-to-noise ratio (PSNR), structural similarity (SSIM), and Frechet inception distance (FID) compared with classical open-loop MISR; difference image block which demonstrates the extraordinary reconstruction capability of the proposed method.

The rest of this paper is organized as follows. Related work is summarized in section 2, the theoretical basis are detailed in section 3, the simulation experiment is designed in section 4, and the final conclusion is drawn in section 5.

## 2 Related Work

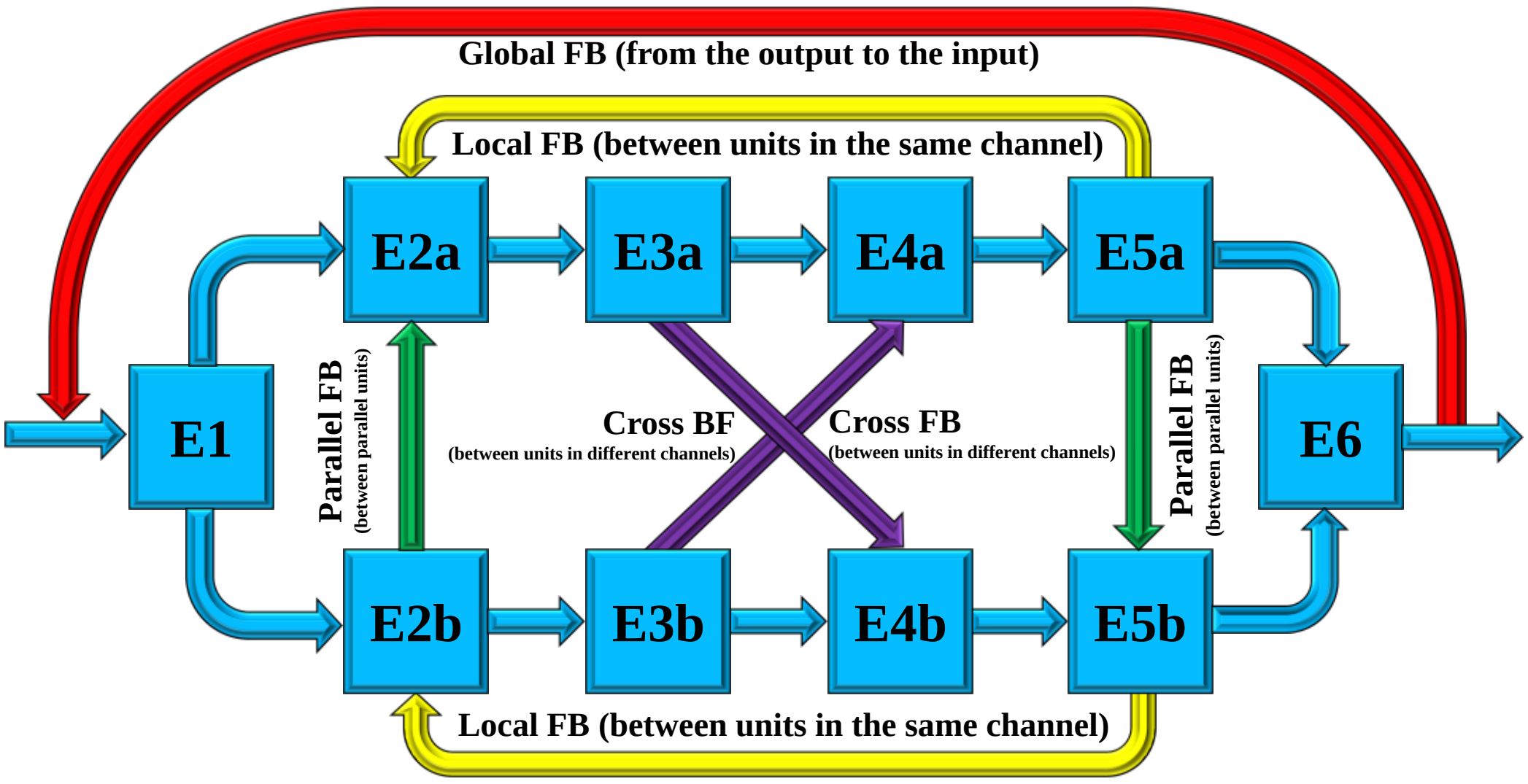

**Figure 1. FB modes in MSIR.**

FB mechanism can markedly improve the performance of MISR by delivering an error term from the front to the end, from the end to the front, from the HQ image to the LQ image, or from the LQ image to the HQ image [25-60]. FB modes in MISR may be partitioned into several categories: local FB and global FB, serial FB and parallel FB, online FB and offline FB, single FB, double FB, and multiple FB, intra-image FB and inter-image FB, and so forth [25-60]. Some FB modes in MISR are displayed in Figure 2 including global FB, local FB, parallel FB, and cross FB, where E is the component of MISR. Some of them are special for MISR [26, 34, 38, 43, 45, 48-49, 60], others are special for nature, depth, remote sensing, or hyper-spectrum ISR. Nevertheless, the latter can be revised for MISR via taking into account the LQ property of medical images.

For local and global FB modes, most FB mechanisms in ISR use local FB. Global FB transmits signal from the output to the input and local FB transmits signal between network units in the same network channel. Wang BB et al. propose a global FB from the HQ image to the LQ image [25-27]. Ge YL et al. present a global FB from the output to the input [28].

For serial and parallel FB modes, most FB mechanisms utilize serial FB. Serial FB delivers signal between network units in the same network channel and parallel FB delivers signal between parallel network units in different network channels. Wang BB et al. also raise lightweight parallel FB among different iterations [25]. Liu XB et al. put forward self-attention negative FB networks with parallel FB modules [29]. Chen X et al. come up with an attention mechanism FB network with parallel FB among iterations [30].

For online and offline FB modes, the former is implemented in the training stage; the latter is implemented in the testing stage, such as plug-and-play form. Most FB mechanisms make use of offline FB [25-60].

For single, double, and multiple FB modes, the double FB comprises progressive FB, mutual FB, cross FB, paired FB, and coupled FB; the multiple FB comprises overlap FB, multilevel FB, multiscale FB, multiple distillation FB, and pyramid FB. Single FB contains one FB signal, double FB contains two FB signals, and multiple FB contains three or more FB signals. Wang BB et al. further propose a dual mutual FB between high-resolution (HR) images and low-resolution images for MISR [26]. Wang BB et al. additionally present a cross-learning FB block between HR and LR images [28, 30]. Dong WQ et al. raise a two-branch cross-FB dense network with context-aware guided attention [28]. Wang F et al. put forward a double-branch projection FB network [33]. Qiu DF et al. come up with a progressive FB with a reconstruction error between HR images and a related reconstruction error between LR images for MRI SR [34]. Wu HL et al. propose a paired FB with saliency and non-saliency maps [35]. Deng X et al. present a deep coupled FB network for a pair of extremely over-exposed and under-exposed LR images [36]. Wang BB et al., moreover, raise an overlapping back-projection FB network to learn the hierarchical representations [37]. Ge YL et al. put forward a multiscale pixel-attention FB link network with a multiscale DS module to reuse the feature information of different depths comprehensively [28]. Ren S et al. bring forward an asymmetric back-projection network with multilevel error FB for medical video SR [38]. Lin ZC et al. come up with an error FB architecture with a dense residual mechanism to fuse multiscale features [39]. Chen X et al. propose a multiscale recursive FB network with multiscale convolution to capture image features of different scales [40]. Xu WJ et al. present a multiscale FB residual network for single ISR [41]. Wang J et al. raise a multiscale FB module that extracts multiscale features and alleviates the information loss in network propagation [42]. Du YB et al. put forward a multiple distillation FB network that iteratively up-samples (US) and down-samples to fully extract the texture details of X-ray images [43]. Wu HP et al. put forward an attention network with pyramid FB to enhance low-level feature expression with high-level information [44]. Shang JR et al. bring forward a gated multi-attention FB network for MISR [45].

For intra-image and inter-image FB modes, inter-image FB is usually adopted in video SR. Intra-image FB sends a signal inside an image and inter-image FB sends a signal between two images. Deng X et al. propose an inter-image FB mechanism for multi-exposure images [36]. Zhu J et al. present a frame-by-frame FB fusion network for video SR [46].

FB signals utilized in ISR contain pixel, edge, error, attention, saliency, and so on [28-30, 32, 34-35, 38, 44, 48-50, 55, 57].

FB operations used in ISR embrace summation, product, concatenation, and so forth [25-60].

FB architectures employed in ISR cover density, convolution, transformer, residual, recurrent, recursion, and generative adversarial network, and so on [32, 34-35, 39-41, 52, 56, 59-60].

It is worth emphasizing that all current FB modes are utilized in the SR unit of an open-loop architecture. This paper will use the FB modes in a closed-loop architecture with UR and SR units to achieve better ISR performance.

All in all, this paper proposes a global, serial, offline, single, and intra-image FB-based closed-loop CMISR.

## 3 Theory

### 3.1 Terminology Glossary

The terminology abbreviations and mathematical notations utilized in this paper are collected in Table 1.

**Table 1. Summary of abbreviations & notations.**

| Abbreviation & Notation | Meaning |
|---|---|
| UR/SR | Under-Resolution/Super-Resolution |
| DS/US | Down-Sampling(Sample)/Up-Sampling(Sample) |
| ISR/MISR/CMISR/LMISR | Image SR/Medical ISR/Circular/Linked MISR |
| HQ/LQ | High-Quality/Low-Quality |
| HR/LR | High-Resolution/Low-Resolution |
| HD/LD | High-Definition/Low-Definition |
| FB/NF | Feedback/Nonlinear Function |
| D/d | HQ Dimension/LQ Dimension |
| $\mathbf{x}/\mathbf{x}_u/\mathbf{x}_s$ | HQ Image/LQ Image/SR Image |
| $\mathbf{x}_e/\mathbf{x}_c$ | Error Term/Control Term |
| **k/n** | Degradation Kernel/Noise Vector |
| **r/R** | First-Order/Higher-Order Item |
| **Λ/U** | Coefficient Matrix |
| t/**e** | Time/Recovery Error |

### 3.2 Theoretical Architecture

The proposed theoretical architecture is illustrated in Figure 2. It consists of two halves: the top half and the bottom half. The top half is the framework of classical MISR. It is an open-loop framework and is named linked, serial, or cascade MISR (LMISR). The bottom half is the closed-loop CMISR framework. LMISR contains two units: UR and SR. The UR unit degrades the HQ image to the LQ image. The SR unit recovers the HR image from the LR image. CMISR incorporates five elements: UR, SR, summator, multiplier, and integrator. The UR and SR elements of CMISR are the same as those of LMISR. The summator introduces negative feedback into CMISR. Architecture input is LQ image $\mathbf{x}_{u0}$ and architecture output is reconstructed HQ image $\mathbf{x}$. The UR unit of LMISR is marked with dotted line because original HQ image $\mathbf{x}_0$ is unknown. CMISR with closed-loop architecture can achieve better HD images from LD images than LMISR.

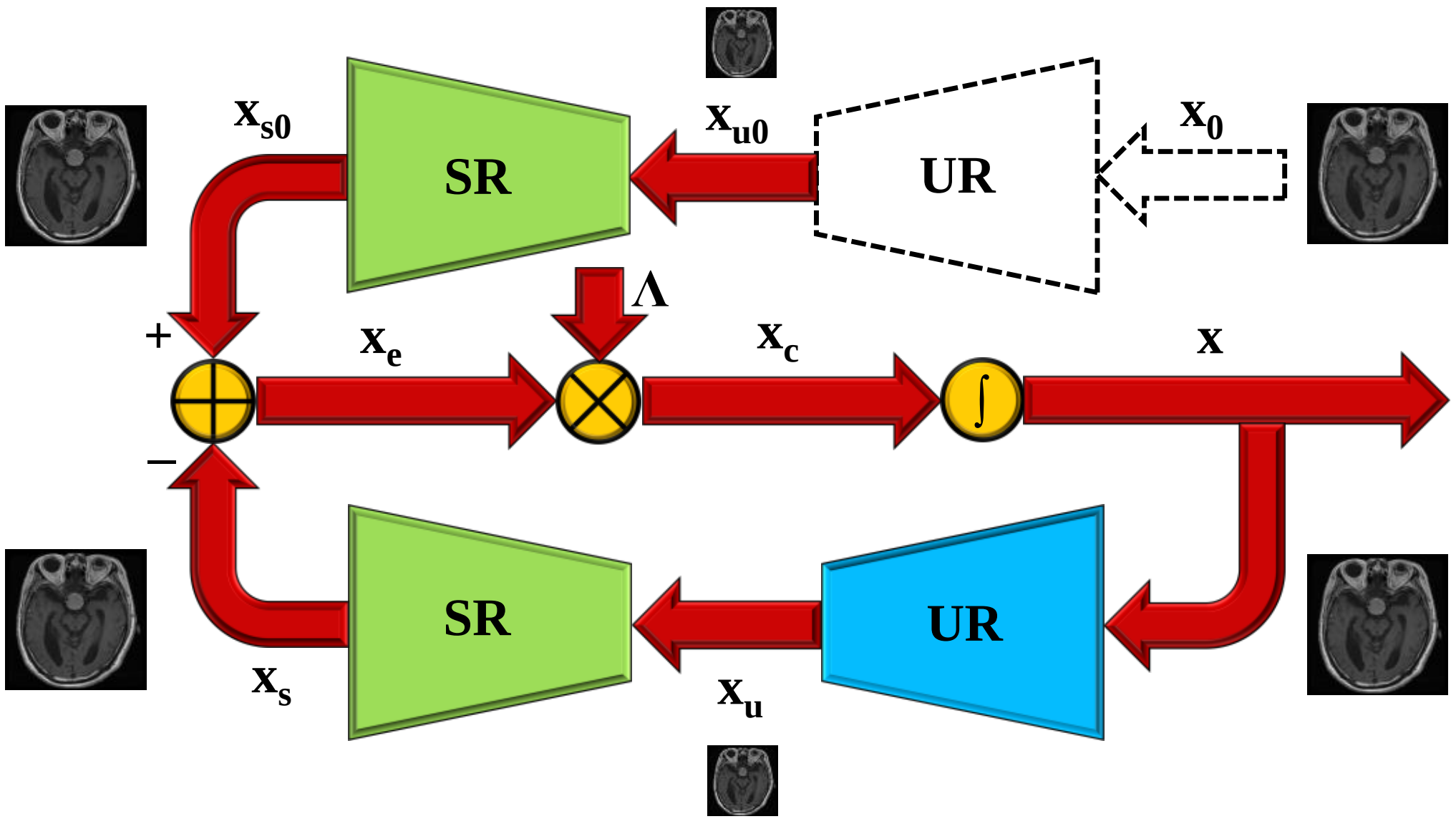

**Figure 2. The theoretical architecture of CMISR.**

The LMISR can be described by the following mathematical equations:

$$\begin{gathered} \mathbf{x}_{u0} = \mathrm{UR}\left(\mathbf{x}_0\right) \\ \mathbf{x}_{s0} = \mathrm{SR}\left(\mathbf{x}_{u0}\right) \\ \mathbf{x}_0, \mathbf{x}_{s0} \in \mathrm{R}^{D}; \mathbf{x}_{u0} \in \mathrm{R}^{d} \end{gathered} , \tag{1}$$

where:
UR is given in classical MISR, such as nearest, bilinear, or bicubic interpolation;
UR is assumed or estimated in blind or real-world MISR, such as a degradation kernel or degradation network mapping from HQ image to LQ image;
SR can be any state-of-the-art ISR approaches including classical, blind, or real-world ISR;
$\mathbf{x}_0$ is the original HQ image, it is actually unknown, and it is introduced for theory analysis;
$\mathbf{x}_{u0}$ is the LQ image after UR and it is the true input of the architecture;
$\mathbf{x}_{s0}$ is the recovered HQ image after SR;
D is the dimension of $\mathbf{x}_0$ and $\mathbf{x}_{s0}$;
d is the dimension of $\mathbf{x}_{u0}$.

LMISR can be formally resolved by the following optimization problem:

$$\begin{aligned} &\mathbf{x}_{s0} = \arg\min_{\mathbf{x}_{s0}} \mathrm{L}\left(\mathbf{x}_{s0}\right) \\ &\text{s.t. } \mathbf{x}_{u0} = \mathrm{UR}\left(\mathbf{x}_0\right), \mathbf{x}_{s0} = \mathrm{SR}\left(\mathbf{x}_{u0}\right), \\ &\mathbf{x}_0, \mathbf{x}_{s0} \in \mathrm{R}^{\mathrm{D}}; \mathbf{x}_{u0} \in \mathrm{R}^{\mathrm{d}} \end{aligned} \tag{2}$$

where: L represents the loss function which is related to model-based or data-driven priors and possible regularization constraints.

For classical LMISR, the UR unit is a given DS module, which can be depicted by the following mathematical expressions:

$$\mathbf{x}_{u0} = \mathrm{UR}\left(\mathbf{x}_0\right) = \mathrm{DS}\left(\mathbf{x}_0\right) = \left(\mathbf{x}_0\right)\downarrow_{s}, \tag{3}$$

where:
↓ denotes DS operation, such as nearest, bilinear, or bicubic interpolation;
s denotes the scale factor.

For blind or real-world LMISR, the UR unit comprises degradation and DS subunits, which can be expressed by the following mathematical formulas:

$$\begin{aligned} &\mathbf{x}_{u0} = \mathrm{UR}\left(\mathbf{x}_0\right) = \begin{cases} \mathrm{DS}\left(\mathbf{x}_0 \otimes \mathbf{k}\right) + \mathbf{n} = \left(\mathbf{x}_0 \otimes \mathbf{k}\right)\downarrow_{s} + \mathbf{n} \\ \mathrm{DS}\left(\mathbf{x}_0\right) \otimes \mathbf{k} + \mathbf{n} = \left(\mathbf{x}_0\right)\downarrow_{s} \otimes \mathbf{k} + \mathbf{n} \end{cases}, \\ &\mathbf{k} \in \mathrm{R}^{\mathrm{K}}, \mathbf{n} \in \mathrm{R}^{\mathrm{d}} \end{aligned} \tag{4}$$

where:
$\otimes$ is convolutional operation;
**k** is the degradation kernel which can be assumed or estimated;
**n** is the random noise with the uniform or Gaussian distribution;
K is the dimension of the degradation kernel.

For classical LMISR, UR is only a simple DS operation; for blind or real-world LMISR, we can always suppose that an LQ image is down-sampled from a degraded HQ image or is degraded from a down-sampled HQ image.

According to Equation (1), the nonlinear function (NF) of LMISR from the input to the output can be described by the following mathematical equations:

$$\begin{gathered} \mathbf{x}_{s0} = \mathrm{SR}\left(\mathrm{UR}\left(\mathbf{x}_0\right)\right) = \mathrm{NF}\left(\mathbf{x}_0\right) \\ \mathrm{NF}(\cdot) = \mathrm{SR}\left(\mathrm{UR}(\cdot)\right) \\ \mathrm{NF}(\cdot) \in \mathrm{R}^{\mathrm{D}} \end{gathered} . \tag{5}$$

The CMISR can be depicted by the following mathematical expressions:

$$\begin{gathered} \mathbf{x}_u(t) = \mathrm{UR}\left(\mathbf{x}(t)\right) \\ \mathbf{x}_s(t) = \mathrm{SR}\left(\mathbf{x}_u(t)\right) \\ \mathbf{x}_e(t) = \mathbf{x}_{s0} - \mathbf{x}_s(t) \\ \mathbf{x}_c(t) = \mathbf{\Lambda}(t)\mathbf{x}_e(t) \\ \mathbf{x}(t) = \mathbf{x}(0) + \int_0^t \mathbf{x}_c(t)\,dt \\ \mathbf{\Lambda}(t) = \mathrm{diag}\left(\mathbf{\Lambda}_{11}(t) \quad \mathbf{\Lambda}_{22}(t) \quad \cdots \quad \mathbf{\Lambda}_{DD}(t)\right) \\ t \in \mathrm{R}; \mathbf{x}(t), \mathbf{x}_s(t), \mathbf{x}_e(t), \mathbf{x}_c(t) \in \mathrm{R}^{\mathrm{D}}; \mathbf{x}_u(t) \in \mathrm{R}^{\mathrm{d}}; \mathbf{\Lambda}(t) \in \mathrm{R}^{\mathrm{D}\times\mathrm{D}} \end{gathered} , \tag{6}$$

where:
t represents the time;
$\mathbf{x}(t)$ represents the expected HR image at time t and it is the output of the architecture;
$\mathbf{x}_u(t)$ represents the LR image of $\mathbf{x}(t)$ at time t;
$\mathbf{x}_s(t)$ represents the recovered HR image of $\mathbf{x}_u(t)$ at time t and it is the output of the architecture;
$\mathbf{x}_e(t)$ represents the error term at time t;
$\mathbf{x}_c(t)$ represents the control term at time t;
$\mathbf{x}(0)$ represents the initial value of $\mathbf{x}(t)$ at time 0 and equals zero vector, random vector, or $\mathbf{x}_0$;
$\mathbf{\Lambda}(t)$ represents the diagonal matrix of multiplication coefficient at time t;
$\mathbf{\Lambda}_{ii}(t)$ represents the i-th principal diagonal element of $\mathbf{\Lambda}(t)$;
diag(·) represents a diagonal matrix.

According to Equations (5) and (6), the closed-loop equation can be expressed by the following mathematical formulas:

$$\begin{gathered} \mathbf{x}(t) = \mathbf{x}(0) + \int_0^t \mathbf{\Lambda}(t)\left(\mathbf{x}_{s0} - \mathrm{SR}\left(\mathrm{UR}\left(\mathbf{x}(t)\right)\right)\right)dt \\ \mathbf{x}(t) = \mathbf{x}(0) + \int_0^t \mathbf{\Lambda}(t)\left(\mathbf{x}_{s0} - \mathrm{NF}\left(\mathbf{x}(t)\right)\right)dt \end{gathered} , \tag{7}$$

### 3.3 Nonlinear Function

The nonlinear function NF(x(t)) in Equation (7) can be expanded with the Taylor series at $\mathbf{x}_0$ by the following mathematical expressions:

$$
\begin{gathered}
\mathrm{NF}(\mathbf{x}(t)) = \mathrm{NF}(\mathbf{x}_0) + \mathbf{r}(t) + \mathbf{R}(t) \\
\mathbf{r}_i(t) = \sum_{j=1}^{D} \left(\mathbf{x}_j(t) - \mathbf{x}_{0j}\right) \left.\frac{\partial \mathrm{NF}_i(\mathbf{x}(t))}{\partial \mathbf{x}_j(t)}\right|_{\mathbf{x}(t)=\mathbf{x}_0} , \\
\mathbf{r}(t), \mathbf{R}(t) \in R^D; i,j = 1,2,\cdots,D
\end{gathered}
\tag{8}
$$

where:

$\mathrm{NF}(\mathbf{x}_0)$ denotes the constant term at $\mathbf{x}_0$;

**r**(t) denotes the first-order term at time t;

**R**(t) denotes the second-order and higher-order term at time t;

$\mathbf{r}_i(t)$ denotes the i-th element of **r**(t);

$\mathbf{x}_j(t)$ denotes the j-th element of **x**(t);

$\mathbf{x}_{0j}$ denotes the j-th element of $\mathbf{x}_0$;

$\mathrm{NF}_i(\mathbf{x}(t))$ denotes the i-th element of NF(**x**(t)).

For the convenience of later theoretical analysis, after discarding the second-order and higher-order term **R**(t), the NF NF(**x**(t)) can be approximated by the following linear function:

$$
\mathrm{NF}(\mathbf{x}(t)) \approx \mathrm{NF}(\mathbf{x}_0) + \mathbf{r}(t) , \tag{9}
$$

The linear term $\mathbf{r}_i(t)$ can be further approximated by the following mathematical formulas:

$$
\begin{gathered}
\mathbf{r}_i(t) \approx \left(\mathbf{x}_i(t) - \mathbf{x}_{0i}\right) \left.\frac{\partial \mathrm{NF}_i(\mathbf{x})}{\partial \mathbf{x}_i(t)}\right|_{\mathbf{x}_i(t)=\mathbf{x}_{0i}} . \\
i = 1,2,\cdots,D
\end{gathered}
\tag{10}
$$

According to Equation (9), the linear term **r**(t) can be rewritten by the following equations:

$$
\begin{gathered}
\mathbf{r}(t) \approx \mathbf{U}(t)\left(\mathbf{x}(t) - \mathbf{x}_0\right) \\
\mathbf{U}(t) = \mathrm{diag}\left(\mathbf{U}_{11}(t) \quad \mathbf{U}_{22}(t) \quad \cdots \quad \mathbf{U}_{DD}(t)\right) \\
\mathbf{U}_{ii}(t) = \left.\frac{\partial \mathrm{NF}_i(\mathbf{x}(t))}{\partial \mathbf{x}_i(t)}\right|_{\mathbf{x}(t)=\mathbf{x}_0} , \\
\mathbf{U}(t) \in R^{D\times D}; i = 1,2,\cdots,D
\end{gathered}
\tag{11}
$$

where:

$\mathbf{U}(t)$ is the coefficient matrix at time t;
$\mathbf{U}_{ii}(t)$ is the i-th principal diagonal element of $\mathbf{U}(t)$.

### 3.4 Recovery Error

The recovery error between the expected HD image and the original HD image can be described by the following mathematical equation:

$$\begin{gathered}\mathbf{e}(t)=\mathbf{x}(t)-\mathbf{x}_0\\ \mathbf{e}(t)\in R^D\end{gathered}. \tag{12}$$

According to Equation (7), the expected HD image at time t+Δt can be expressed by the following mathematical equation:

$$\begin{gathered}\mathbf{x}(t+\Delta t)=\mathbf{x}(t)+\int_t^{t+\Delta t}\mathbf{\Lambda}(t)\left(\mathbf{x}_{s0}-NF(\mathbf{x}(t))\right)dt\\ \text{s.t. } \Delta t>0\end{gathered}. \tag{13}$$

Subtracting $\mathbf{x}_0$ from both sides of Equation (13), the following mathematical equation can be obtained:

$$\mathbf{x}(t+\Delta t)-\mathbf{x}_0=\mathbf{x}(t)-\mathbf{x}_0+\int_t^{t+\Delta t}\mathbf{\Lambda}(t)\left(\mathbf{x}_{s0}-NF(\mathbf{x}(t))\right)dt. \tag{14}$$

According to Equation (12), the following mathematical expressions can be gained:

$$\begin{gathered}\mathbf{e}(t+\Delta t)=\mathbf{e}(t)+\int_t^{t+\Delta t}\mathbf{\Lambda}(t)\left(\mathbf{x}_{s0}-NF(\mathbf{x}(t))\right)dt\\ \mathbf{e}(t+\Delta t)=\mathbf{x}(t+\Delta t)-\mathbf{x}_0\end{gathered}, \tag{15}$$

where: $\mathbf{e}(t+\Delta t)$ represents the error vector at time t+Δt.
If Δt is close to zero, Equation (15) can be approximated by the following mathematical formula:

$$\begin{gathered}\mathbf{e}(t+\Delta t)\approx\mathbf{e}(t)+\mathbf{\Lambda}(t)\left(\mathbf{x}_{s0}-NF(\mathbf{x}(t))\right)\Delta t\\ \text{s.t. } \Delta t\to 0\end{gathered}. \tag{16}$$

According to (5), (9), and (11), the following mathematical equations can be acquired:

$$\begin{aligned}\mathbf{e}(t+\Delta t)&\approx\mathbf{e}(t)-\mathbf{\Lambda}(t)\mathbf{r}(t)\Delta t\\ &\approx\mathbf{e}(t)-\mathbf{\Lambda}(t)\mathbf{U}(t)\left(\mathbf{x}(t)-\mathbf{x}_0\right)\Delta t\\ &=\mathbf{e}(t)-\mathbf{\Lambda}(t)\mathbf{U}(t)\mathbf{e}(t)\Delta t\\ &=\left(\mathbf{I}-\mathbf{\Lambda}(t)\mathbf{U}(t)\Delta t\right)\mathbf{e}(t)\\ &=\left(\mathbf{I}-\Delta t\mathbf{\Lambda}(t)\mathbf{U}(t)\right)\mathbf{e}(t)\end{aligned}, \tag{17}$$

where **I** denotes the unit matrix.

Computing 2-norm in both sides of Equation (17), the following mathematical inequation can be obtained:

$$\left\|\mathbf{e}(t+\Delta t)\right\|_2 \le \left\|\mathbf{I}-\Delta t\mathbf{\Lambda}(t)\mathbf{U}(t)\right\|_F \cdot \left\|\mathbf{e}(t)\right\|_2, \tag{18}$$

where:
subscript 2 means 2-norm;
subscript F means Frobenius-norm.

We can always choose suitable **Λ**(t) and **U**(t) to satisfy the following inequality:

$$\left\|\mathbf{I}-\Delta t\mathbf{\Lambda}(t)\mathbf{U}(t)\right\|_F < 1. \tag{19}$$

For example, **Λ**(t) and **U**(t) are respectively proportional to a unit matrix:

$$\begin{aligned} &\mathbf{\Lambda}(t)=\lambda\mathbf{I} \\ &\mathbf{U}(t)=\mu(t)\mathbf{I} \\ &\mu(t)=\frac{1}{D}\sum_{i=1}^{D}\mathbf{U}_{ii}(t)=\frac{1}{D}\sum_{i=1}^{D}\left.\frac{\partial NF_i(\mathbf{x}(t))}{\partial \mathbf{x}_i(t)}\right|_{\mathbf{x}(t)=\mathbf{x}_0} \end{aligned}. \tag{20}$$

where:
λ is a constant;
$\mu(t)$ is the average of $\mathbf{U}_{ii}(t)$ in Equation (11).

According to Equation (20), we can always find a proper λ to meet Inequation (19):

$$\begin{aligned} &\left\|\mathbf{I}-\Delta t\mathbf{\Lambda}(t)\mathbf{U}(t)\right\|_F=\left\|\mathbf{I}-\Delta t\cdot\lambda\mathbf{I}\cdot u(t)\mathbf{I}\right\|_F=\left\|(1-\Delta t\cdot\lambda\cdot u(t))\mathbf{I}\right\|_F \\ &=\left|1-\Delta t\cdot u(t)\cdot\lambda\right|\cdot\left\|\mathbf{I}\right\|_F=\left|1-\Delta t\cdot u(t)\cdot\lambda\right|\cdot\sqrt{D}<1 \\ &\Rightarrow\begin{cases} \dfrac{1-\frac{1}{\sqrt{D}}}{\Delta t\cdot u(t)}<\lambda<\dfrac{1+\frac{1}{\sqrt{D}}}{\Delta t\cdot u(t)}, u(t)>0 \\ \dfrac{1+\frac{1}{\sqrt{D}}}{\Delta t\cdot u(t)}<\lambda<\dfrac{1-\frac{1}{\sqrt{D}}}{\Delta t\cdot u(t)}, u(t)<0 \end{cases} \end{aligned}. \tag{21}$$

According to Equations (18) and (19), the following mathematical inequality can be obtained:

$$\left\|\mathbf{e}(t+\Delta t)\right\|_2 < \left\|\mathbf{e}(t)\right\|_2 . \tag{22}$$

According to Inequation (22), if time t is close to infinite in steady-state, the recovery error approximates to zero, and $\mathbf{x}(t)$ approximates to $\mathbf{x}_0$:

$$\begin{gathered} \mathbf{e}(t) = \mathbf{x}(t) - \mathbf{x}_0 \to 0 \\ \mathbf{x}(t) \to \mathbf{x}_0 \\ \text{s.t. } t \to \infty \end{gathered} . \tag{23}$$

Therefore, the proposed CMISR can achieve perfect HD recovery image $\mathbf{x}(t)$ which is close to the original HD image $\mathbf{x}_0$.

### 3.5 Algorithm Description

The proposed CMISR algorithm is described in Figure 3, where N is the total number of iterations.

```
Algorithm: CMISR
Input: x0
Initialization:
   n = 1, xs0 = NF(x0), x = xs0
while n <= N
   update x according to equation (7), (9), (14) and (20)
   n = n+1
end
Output: x
```

**Figure 3. CMISR algorithm description.**

## 4 Experiment

### 4.1 Experimental Conditions

Experimental medical image datasets include three categories: MRI, CT, and X-ray, which are listed in Table 2 [61-63]. The dataset name, image size, and image number of MRI, CT, and X-ray datasets are gathered in Table 2. For the purpose of simplifying implementation, a partial subset of the original dataset is utilized for all algorithms, and the full dataset is merely utilized for two representative algorithms. A partial subset is randomly chosen from the original dataset. The original images in a dataset are regarded as the HQ images and the related LQ images are generated by down-sampling the HQ images with bicubic interpolation. Although medical images usually have one illumination channel, they are repeated to three channels for increasing the performance of SR reconstruction.

**Table 2. Experimental medical image datasets.**

| Name | Size | Number | | | | | |
|---|---|---|---|---|---|---|---|
| | | Full Datasets | | | Partial Subsets | | |
| | | Total | Finetuning | Testing | Total | Finetuning | Testing |
| MRI [61] | 512×512×1 | 3064 | 2700 | 364 | 766 | 700 | 66 |
| CT [62] | unfixed | 2481 | 2200 | 281 | 1229 | 1100 | 129 |
| X-Ray [63] | 300×300×1 | 10192 | 9000 | 1192 | 1100 | 1000 | 100 |

**Table 3. Pretrained models.**

| Algorithm | Scale | Name |
|---|---|---|
| HAT | ×2 | HAT-L_SRx2_ImageNet-pretrain.pth |
| | ×3 | HAT-L_SRx3_ImageNet-pretrain.pth |
| | ×4 | HAT-L_SRx4_ImageNet-pretrain.pth |
| | Link | https://github.com/XPixelGroup/HAT |
| SRFormer | ×2 | SRFormer_SRx2_DF2K.pth |
| | ×3 | SRFormer_SRx3_DF2K.pth |
| | ×4 | SRFormer_SRx4_DF2K.pth |
| | Link | https://github.com/HVision-NKU/SRFormer |
| DAT | ×2 | DAT_x2.pth |
| | ×3 | DAT_x3.pth |
| | ×4 | DAT_x4.pth |
| | Link | https://github.com/zhengchen1999/DAT |
| EMT | ×2 | EMT_LSR_x2.pth |
| | ×3 | EMT_LSR_x3.pth |
| | ×4 | EMT_LSR_x4.pth |
| | Link | https://github.com/Fried-Rice-Lab/EMT |
| SwinIR | ×2 | 001_classicalSR_DIV2K_s48w8_SwinIR-M_x2.pth |
| | ×3 | 001_classicalSR_DIV2K_s48w8_SwinIR-M_x3.pth |
| | ×4 | 001_classicalSR_DIV2K_s48w8_SwinIR-M_x4.pth |
| | Link | https://github.com/JingyunLiang/SwinIR |

The proposed CMISR methods possess the property of plug-and-play and can be established on any existing advanced ISR approaches. Hence, five CMISR methods, circular HAT (CHAT), circular SRFormer (CSRFormer), Circular DAT (CDAT), Circular EMT (CEMT), and circular SwinIR (CSWinIR), are proposed as representations. CHAT is based on the classical HAT, CSRFormer is on the basis of classical SRFormer, CDAT depends on classical DAT, CEMT relies on classical EMT, and CSwinIR is on the strength of classical SwinIR. For the sake of increasing performance and saving computation resources, pretrained models of HAT, SRFormer, DAT, EMT, and SwinIR are utilized, which are enumerated in Table 3. Because HAT, SRFormer, DAT, EMT, and SwinIR are not used exclusively for MISR but for natural ISR, the pretrained models are finely adjusted by finetuning samples of medical image datasets in Table 2.

Three performance metrics, PSNR, SSIM, and FID, are selected to evaluate the capability of ISR. FID is employed to overcome the disadvantages of the classical PSNR and SSIM. For SR HQ image $\mathbf{x}_s$ and original HQ image $\mathbf{x}_0$, the definitions of PSNR, SSIM, and FID are described by the following mathematical equations:

$$\mathrm{PSNR}\left(\mathbf{x}_s,\mathbf{x}_0\right)=10\lg\left(\frac{255^2}{\frac{1}{D}\sum_{i=1}^{D}\left(\mathbf{x}_{si}-\mathbf{x}_{0i}\right)^2}\right), \tag{24}$$

$$\mathrm{SSIM}\left(\mathbf{x}_s,\mathbf{x}_0\right)=\frac{\left(2\mu_{x_s}\mu_{x_0}+\left(0.01\times255\right)^2\right)\left(2\sigma_{x_s x_0}+\left(0.03\times255\right)^2\right)}{\left(\mu_{x_s}^2+\mu_{x_0}^2+\left(0.01\times255\right)^2\right)\left(\sigma_{x_s}^2+\sigma_{x_0}^2+\left(0.03\times255\right)^2\right)}, \tag{25}$$

$$\mathrm{FID}\left(\left(\mathbf{m}_{x_s},\boldsymbol{\Sigma}_{x_s}\right),\left(\mathbf{m}_{x_0},\boldsymbol{\Sigma}_{x_0}\right)\right)=\sqrt{\left\|\mathbf{m}_{x_s}-\mathbf{m}_{x_0}\right\|^2+\mathrm{Tr}\left(\boldsymbol{\Sigma}_{x_s}+\boldsymbol{\Sigma}_{x_0}-2\sqrt{\boldsymbol{\Sigma}_{x_s}\boldsymbol{\Sigma}_{x_0}}\right)}, \tag{26}$$

where:
$\mathbf{x}_{si}$ is the i-th element of $\mathbf{x}_s$;
$\mathbf{x}_{0i}$ is the i-th element of $\mathbf{x}_0$;
$\mu_{xs}$ is the element-wised mean of $\mathbf{x}_s$;
$\mu_{x0}$ is the element-wised mean of $\mathbf{x}_0$;
$\sigma_{xs}$ is the element-wised standard deviation of $\mathbf{x}_s$;
$\sigma_{x0}$ is the element-wised standard deviation of $\mathbf{x}_0$;
$\sigma_{xsx0}$ is the element-wised covariance of $\mathbf{x}_s$ and $\mathbf{x}_0$;
$\mathbf{m}_{xr}$ is the vector-wised mean of the image set of $\mathbf{x}_s$;
$\mathbf{m}_{x0}$ is the vector-wised mean of the image set of $\mathbf{x}_0$;
$\boldsymbol{\Sigma}_{xr}$ is the covariance matrix of the image set of $\mathbf{x}_r$;
$\boldsymbol{\Sigma}_{x0}$ is the covariance matrix of the image set of $\mathbf{x}_0$;
Tr is the matrix trace or the summation of the diagonal elements of a matrix.

The experimental hardware platforms include Intel CPU and Nvidia GPU. The experimental software platforms include Microsoft COLAB and MathWorks MATLAB running on Windows or Linux operating systems. Detailed hardware and software configurations are listed in Table 4.

**4.2 Experimental Results**

The experimental results of HAT and CHAT on partial datasets are listed in Table 5. It is shown in Table 5 that CHAT dramatically outperforms HAT at three image scales (x2, x3, and x4), on three partial datasets, and in PSNR, SSIM, and FID. The experimental results of SRFormer and CSRFormer on partial datasets are enumerated in Table 6. It is manifested in Table 6 that CSRFormer observably overmatches SRFormer at three scales, on three partial datasets, and in PSNR, SSIM and FID. The experimental results of DAT and CDAT on partial datasets are displayed in Table 7. It is indicated in Table 7 that CDAT obviously outbalances DAT at three scales, on three partial datasets, and in PSNR, SSIM and FID. The experimental results of EMT and CEMT on partial datasets are exhibited in Table 8. It is revealed in Table 8 that CEMT distinctly surpasses EMT at three scales, on three partial datasets, and in PSNR, SSIM, and FID. The experimental results of SwinIR and CSwinIR on partial datasets are demonstrated in Table 9. It is uncovered in Table 9 that CSwinIR is significantly superior to SwinIR at three scales, on three partial datasets, and in PSNR, SSIM, and FID. It is also shown in Tables 5 to 9 that PSNR and SSIM decrease and FID increases while scale increases. In a word, the proposed closed-loop CMISR method outperforms the state-of-the-art open-loop ISR in the reconstruction capacity.

**Table 4. Detailed hardware and software configurations.**

| Hardware Configurations | |
|---|---|
| Item | Value |
| CPU Type | Intel Core i7 |
| CPU Memory | 8GB |
| GPU Type | NVIDIA Tesla V100 |
| GPU Memory | 16GB |
| Software Configurations | |
| Item | Value |
| Learning Rate | 0.00001 |
| Total Iterations | 20000 |
| Input/Output Channels | 3/3 |
| Batch Size | 10 |
| Block Size | 64×64 |

**Table 5. Experimental results of HAT and CHAT on partial datasets.**

| Scale | Partial Datasets | PSNR (dB) ↑ | | SSIM ↑ | | FID ↓ | |
|---|---|---|---|---|---|---|---|
| | | HAT | CHAT | HAT | CHAT | HAT | CHAT |
| ×2 | MRI | 50.0248 | **50.5429** | 0.9958 | **0.9960** | 0.0051 | **0.0046** |
| | CT | 42.5318 | **43.4011** | 0.9750 | **0.9779** | 0.0086 | **0.0078** |
| | X-Ray | 40.7668 | **41.7838** | 0.9885 | **0.9891** | 0.0128 | **0.0126** |
| ×3 | MRI | 42.1251 | **42.6183** | 0.9831 | **0.9840** | 0.0105 | **0.0102** |
| | CT | 37.7106 | **38.2736** | 0.9330 | **0.9369** | 0.0174 | **0.0152** |
| | X-Ray | 37.8307 | **39.5381** | 0.9643 | **0.9688** | 0.0235 | **0.0234** |
| ×4 | MRI | 38.3789 | **38.8276** | 0.9650 | **0.9668** | 0.0144 | **0.0138** |
| | CT | 35.0430 | **35.9115** | 0.9033 | **0.9100** | 0.0237 | **0.0224** |
| | X-Ray | 35.6091 | **37.3330** | 0.9310 | **0.9438** | 0.0311 | **0.0289** |

**Table 6. Experimental results of SRFormer and CSRFormer on partial datasets.**

| Scale | Partial Datasets | PSNR (dB) ↑ | | SSIM ↑ | | FID ↓ | |
|---|---|---|---|---|---|---|---|
| | | SRFormer | CSRFormer | SRFormer | CSRFormer | SRFormer | CSRFormer |
| ×2 | MRI | 49.9805 | **50.6102** | 0.9958 | **0.9962** | 0.0053 | **0.0046** |
| | CT | 41.6263 | **42.1799** | 0.9717 | **0.9733** | 0.0095 | **0.0088** |
| | X-Ray | 41.1794 | **43.3741** | 0.9883 | **0.9890** | 0.0132 | **0.0127** |
| ×3 | MRI | 41.9589 | **42.4712** | 0.9828 | **0.9839** | 0.0111 | **0.0100** |
| | CT | 37.2787 | **38.0911** | 0.9312 | **0.9359** | 0.0182 | **0.0154** |
| | X-Ray | 37.4805 | **39.0765** | 0.9621 | **0.9679** | 0.0261 | **0.0217** |
| ×4 | MRI | 38.2340 | **38.6690** | 0.9645 | **0.9663** | 0.0150 | **0.0140** |
| | CT | 34.2941 | **35.7712** | 0.9000 | **0.9089** | 0.0246 | **0.0225** |
| | X-Ray | 35.1524 | **36.9347** | 0.9263 | **0.9423** | 0.0322 | **0.0289** |

**Table 7. Experimental results of DAT and CDAT on partial datasets.**

| Scale | Partial Datasets | PSNR (dB) ↑ | | SSIM ↑ | | FID ↓ | |
|---|---|---|---|---|---|---|---|
| | | DAT | CDAT | DAT | CDAT | DAT | CDAT |
| ×2 | MRI | 49.9177 | **50.7162** | 0.9958 | **0.9963** | 0.0047 | **0.0044** |
| | CT | 41.8785 | **42.4962** | 0.9720 | **0.9742** | 0.0092 | **0.0088** |
| | X-Ray | 44.0272 | **45.0623** | 0.9886 | **0.9891** | 0.0128 | **0.0111** |
| ×3 | MRI | 42.0649 | **42.5741** | 0.9830 | **0.9840** | 0.0111 | **0.0104** |
| | CT | 37.6365 | **38.2154** | 0.9320 | **0.9363** | 0.0174 | **0.0154** |
| | X-Ray | 38.7469 | **40.3096** | 0.9631 | **0.9686** | 0.0232 | **0.0191** |
| ×4 | MRI | 38.2398 | **38.7252** | 0.9646 | **0.9666** | 0.0149 | **0.0142** |
| | CT | 34.4314 | **35.9167** | 0.8996 | **0.9097** | 0.0249 | **0.0239** |
| | X-Ray | 35.5294 | **37.5560** | 0.9271 | **0.9436** | 0.0312 | **0.0250** |

**Table 8. Experimental results of EMT and CEMT on partial datasets.**

| Scale | Partial Datasets | PSNR (dB) ↑ | | SSIM ↑ | | FID ↓ | |
|---|---|---|---|---|---|---|---|
| | | EMT | CEMT | EMT | CEMT | EMT | CEMT |
| ×2 | MRI | 49.8131 | **50.0689** | 0.9956 | **0.9959** | 0.0048 | **0.0046** |
| | CT | 41.7036 | **41.8523** | 0.9712 | **0.9719** | 0.0093 | **0.0092** |
| | X-Ray | 44.3858 | **44.8477** | 0.9878 | **0.9885** | 0.0122 | **0.0110** |
| ×3 | MRI | 41.8422 | **42.2202** | 0.9825 | **0.9832** | 0.0116 | **0.0107** |
| | CT | 37.3866 | **38.0442** | 0.9305 | **0.9349** | 0.0181 | **0.0161** |
| | X-Ray | 38.2052 | **38.7967** | 0.9596 | **0.9635** | 0.0304 | **0.0229** |
| ×4 | MRI | 38.1516 | **38.5168** | 0.9640 | **0.9655** | 0.0144 | **0.0139** |
| | CT | 34.2089 | **35.7463** | 0.8991 | **0.9082** | 0.0258 | **0.0217** |
| | X-Ray | 35.2043 | **37.0394** | 0.9232 | **0.9402** | 0.0341 | **0.0277** |

**Table 9. Experimental results of SwinIR and CSwinIR on partial datasets.**

| Scale | Partial Datasets | PSNR (dB) ↑ | | SSIM ↑ | | FID ↓ | |
|---|---|---|---|---|---|---|---|
| | | SwinIR | CSwinIR | SwinIR | CSwinIR | SwinIR | CSwinIR |
| ×2 | MRI | 49.7368 | **49.7958** | 0.9957 | **0.9957** | 0.0050 | **0.0050** |
| | CT | 41.4035 | **42.0304** | 0.9698 | **0.9736** | 0.0097 | **0.0094** |
| | X-Ray | 42.8964 | **42.9702** | 0.9880 | **0.9880** | 0.0123 | **0.0120** |
| ×3 | MRI | 41.8210 | **41.8758** | 0.9825 | **0.9826** | 0.0108 | **0.0107** |
| | CT | 37.1658 | **37.2217** | 0.9291 | **0.9293** | 0.0181 | **0.0180** |
| | X-Ray | 38.2099 | **38.7434** | 0.9602 | **0.9654** | 0.0270 | **0.0217** |
| ×4 | MRI | 38.1044 | **38.1341** | 0.9639 | **0.9640** | 0.0182 | **0.0138** |
| | CT | 34.1425 | **34.2244** | 0.8981 | **0.8981** | 0.0241 | **0.0239** |
| | X-Ray | 35.2544 | **36.4657** | 0.9242 | **0.9371** | 0.0363 | **0.0267** |

**Table 10. Experimental results of HAT and CHAT on full datasets.**

| Scale | Full Datasets | PSNR (dB) ↑ | | SSIM ↑ | | FID ↓ | |
|---|---|---|---|---|---|---|---|
| | | HAT | CHAT | HAT | CHAT | HAT | CHAT |
| ×2 | MRI | 50.2930 | **50.7830** | 0.9953 | **0.9963** | 0.0047 | **0.0041** |
| | CT | 43.1129 | **43.4490** | 0.9792 | **0.9801** | 0.0079 | **0.0074** |
| | X-Ray | 40.6869 | **43.7794** | 0.9908 | **0.9916** | 0.0163 | **0.0160** |
| ×3 | MRI | 42.0737 | **42.1177** | 0.9803 | **0.9803** | 0.0115 | **0.0099** |
| | CT | 38.0818 | **38.1220** | 0.9429 | **0.9429** | 0.0158 | **0.0144** |
| | X-Ray | 39.5443 | **41.5941** | 0.9765 | **0.9776** | 0.0308 | **0.0273** |
| ×4 | MRI | 37.5794 | **37.6670** | 0.9538 | **0.9559** | 0.0167 | **0.0157** |
| | CT | 35.3299 | **35.3940** | 0.9140 | **0.9140** | 0.0223 | **0.0205** |
| | X-Ray | 38.3154 | **38.8848** | 0.9579 | **0.9604** | 0.0378 | **0.0368** |

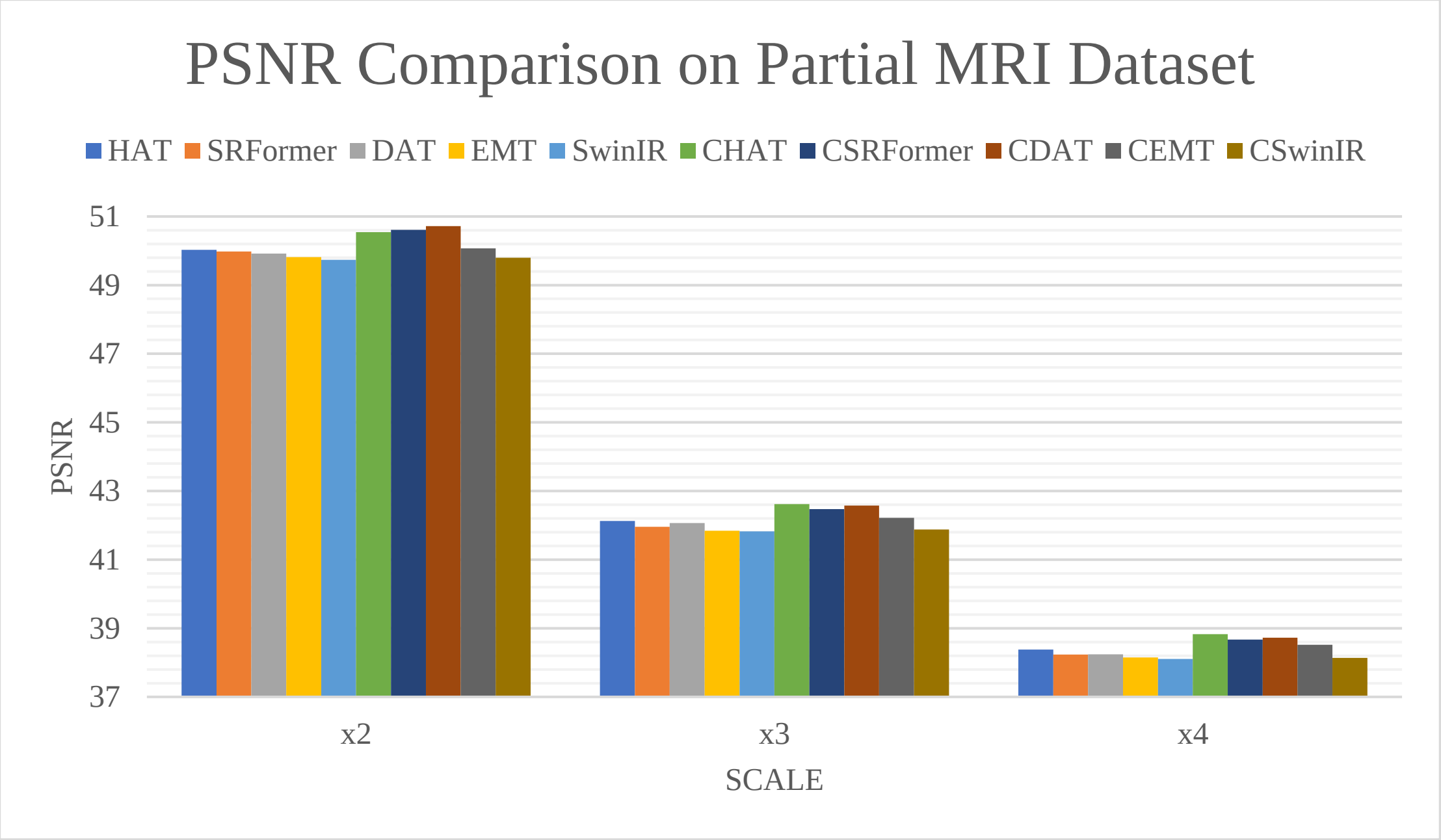

**Figure 4. PSNR comparison on partial MRI dataset.**

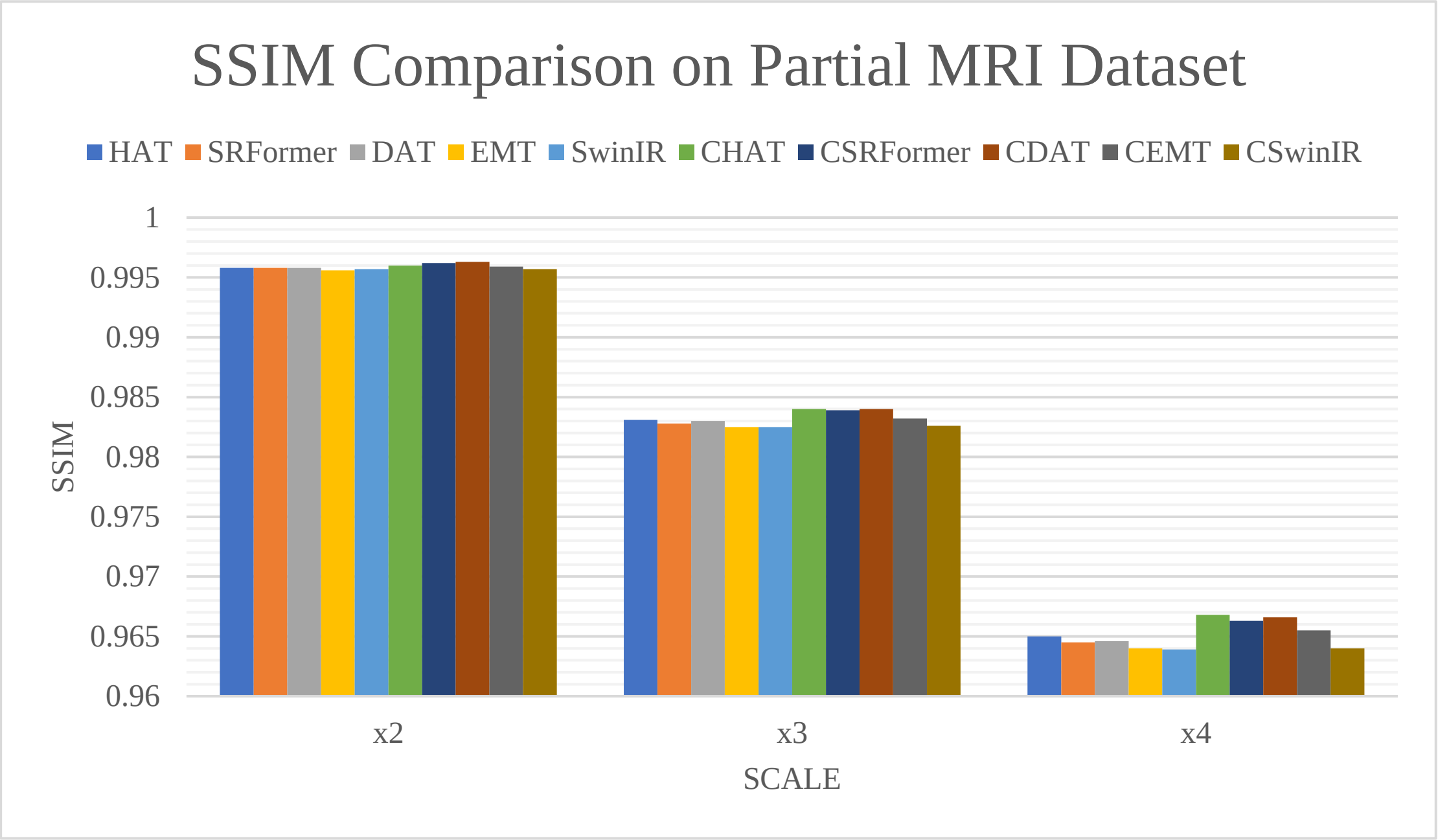

**Figure 5. SSIM comparison on partial MRI dataset.**

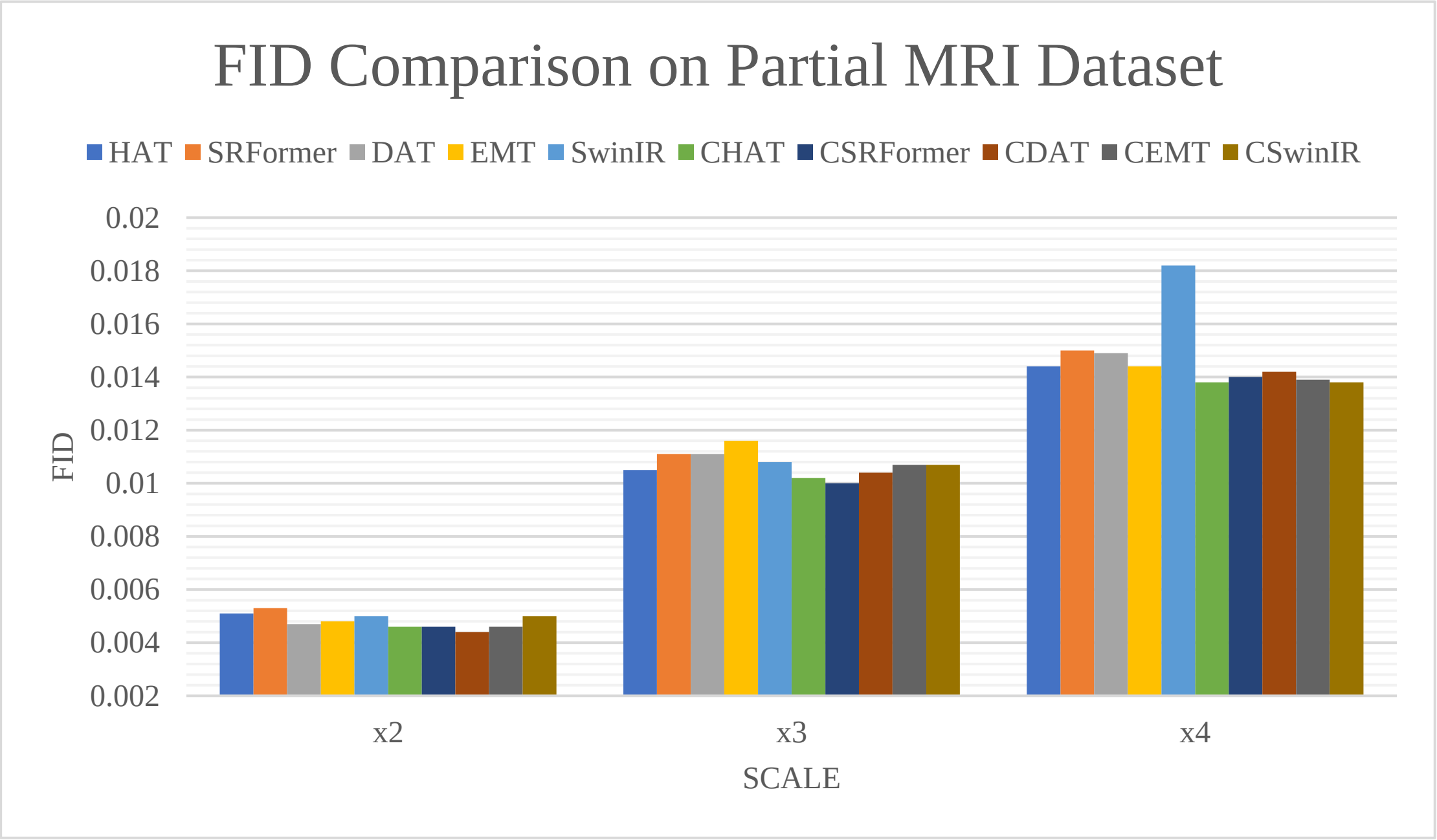

**Figure 6. FID comparison on partial MRI dataset.**

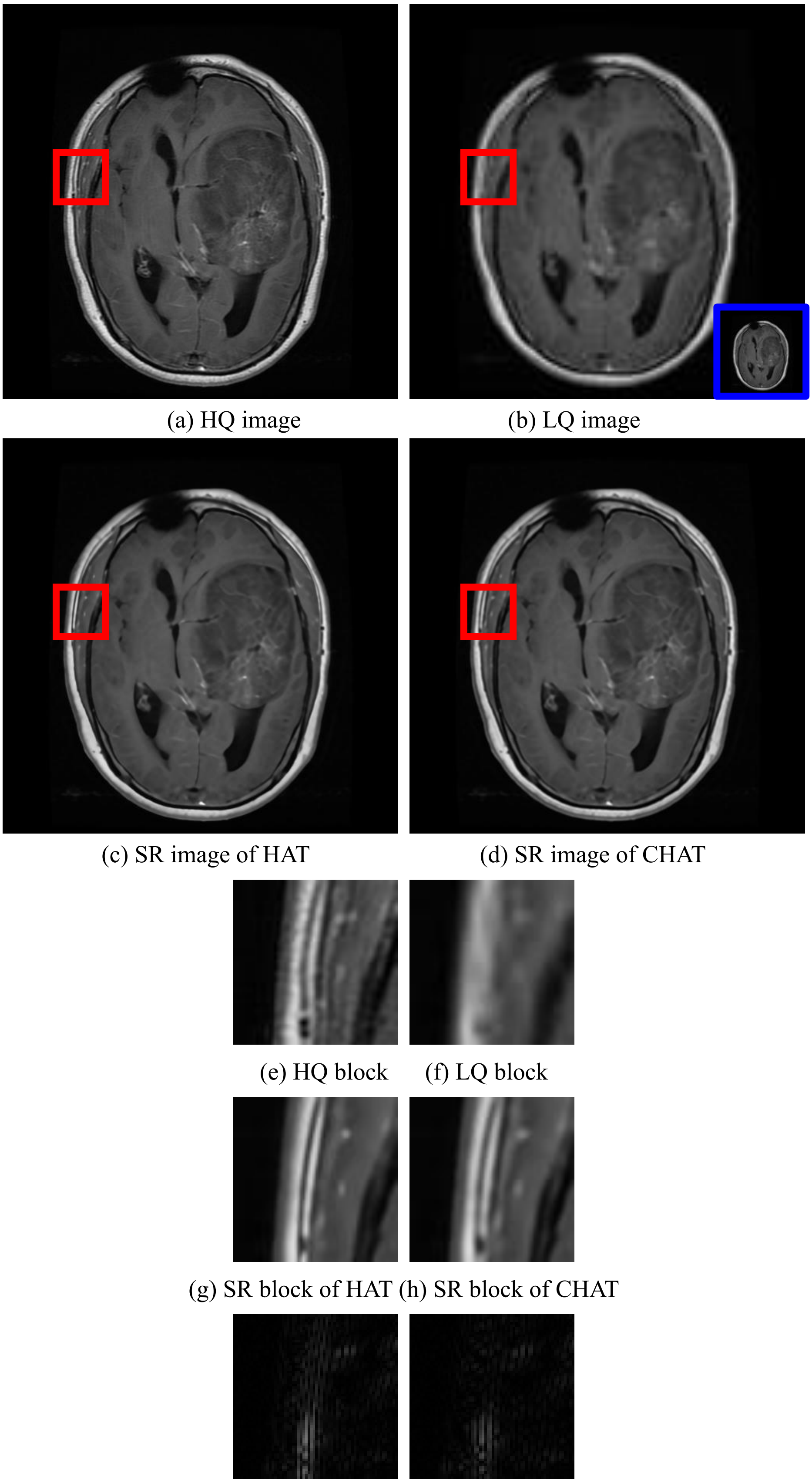

(a) HQ image (b) LQ image

(c) SR image of HAT (d) SR image of CHAT

(e) HQ block (f) LQ block

(g) SR block of HAT (h) SR block of CHAT

(i) difference block of HAT (j) difference block of CHAT

**Figure 7. Experimental results of HAT and CHAT at scale ×4 on partial MRI dataset.**

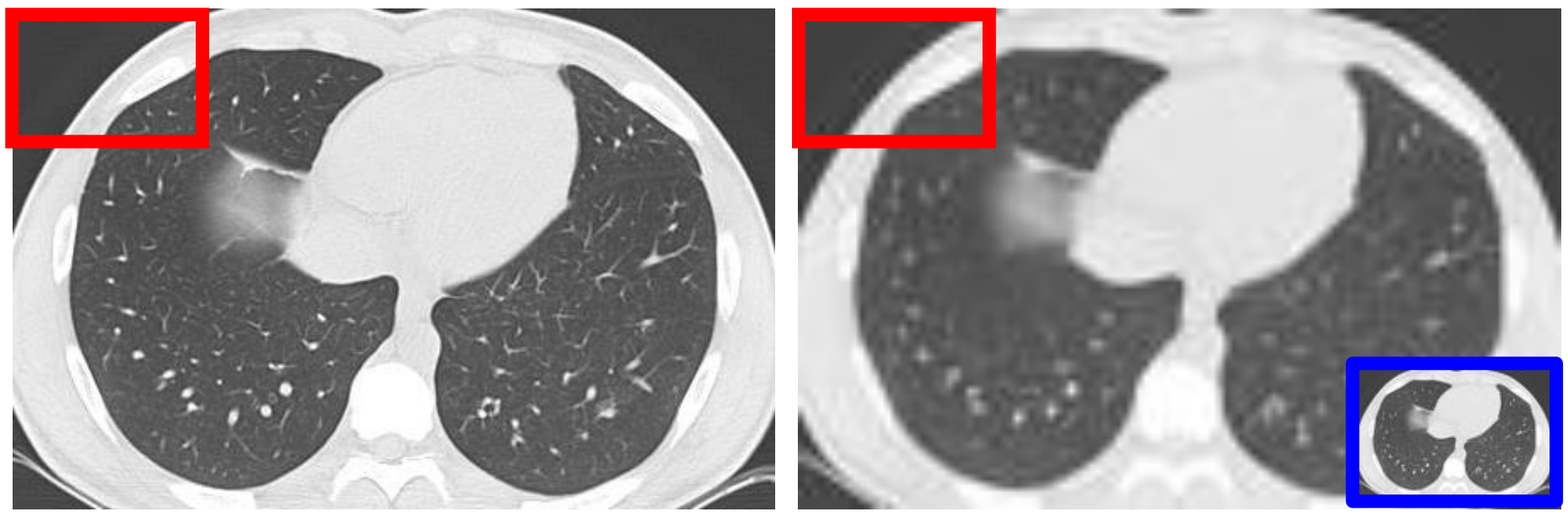

(a) HQ image (b) LQ image

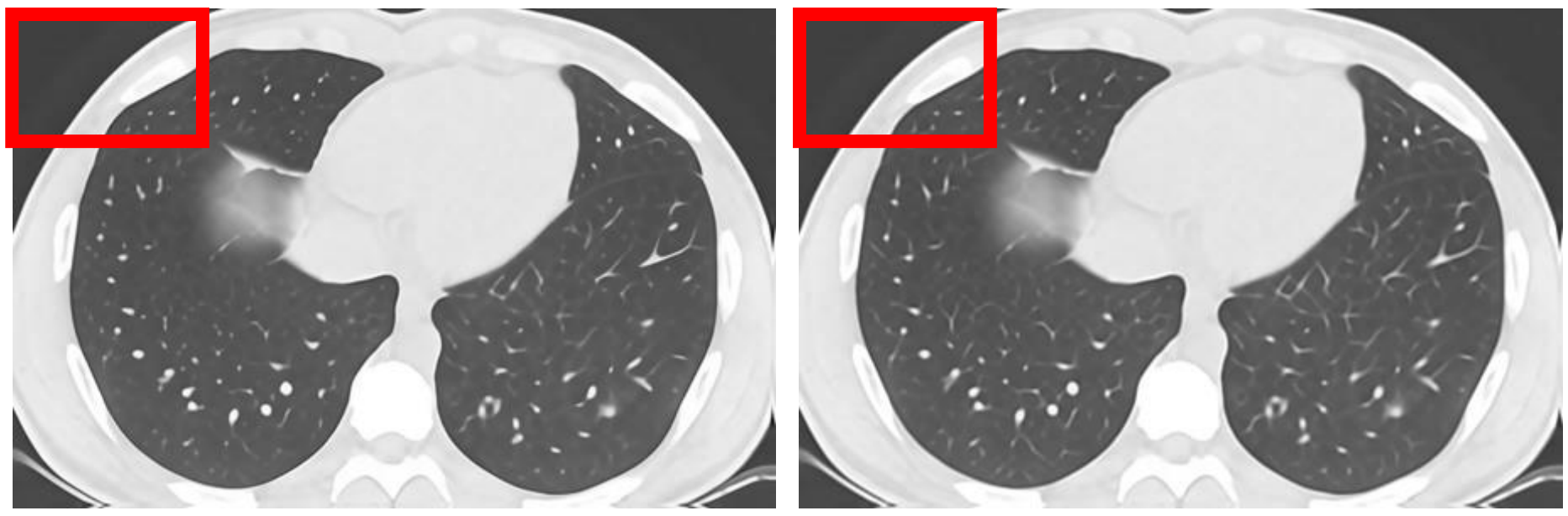

(c) SR image of HAT (d) SR image of CHAT

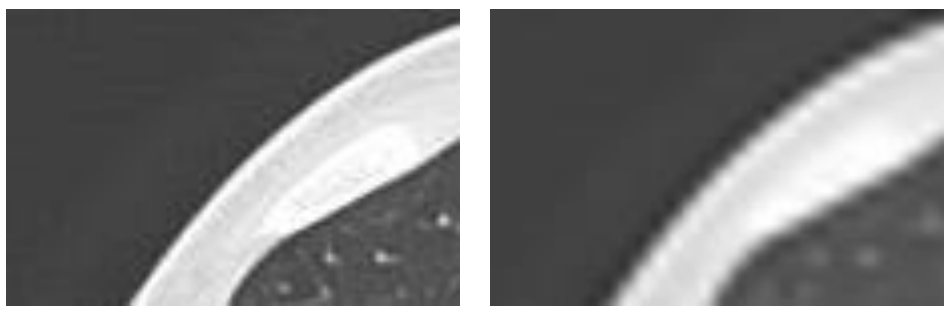

(e) HQ block (f) LQ block

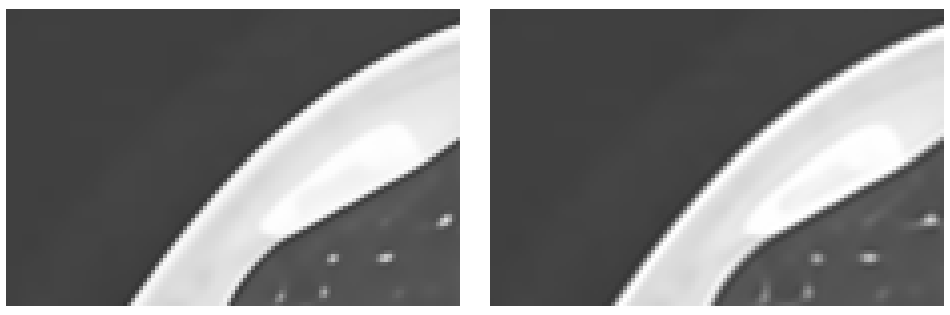

(g) SR block of HAT (h) SR block of CHAT

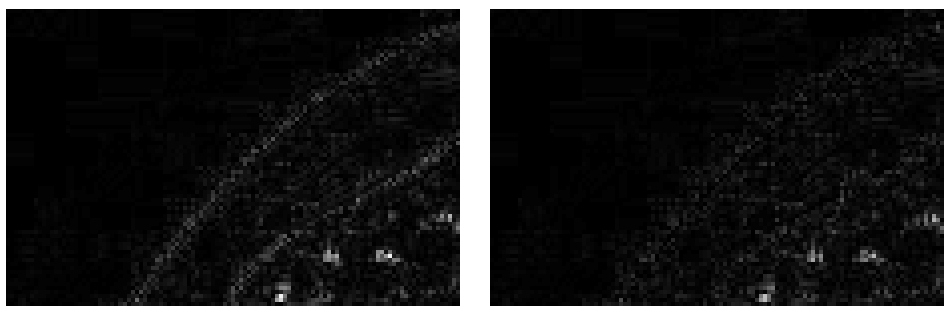

(i) difference block of HAT (j) difference block of CHAT

**Figure 8. Experimental results of HAT and CHAT at scale ×4 on partial CT dataset.**

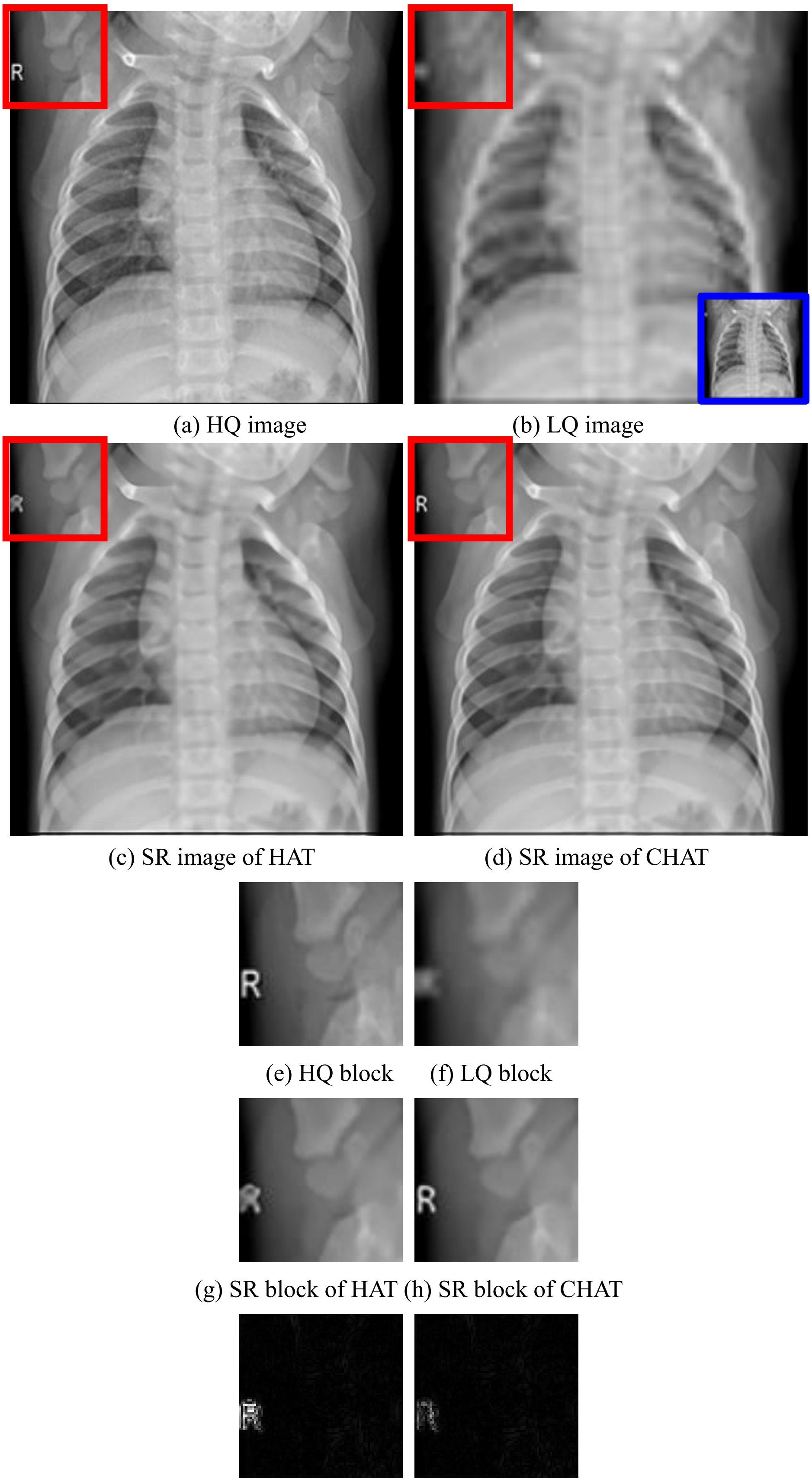

(a) HQ image (b) LQ image

(c) SR image of HAT (d) SR image of CHAT

(e) HQ block (f) LQ block

(g) SR block of HAT (h) SR block of CHAT

(i) difference block of HAT (j) difference block of CHAT

**Figure 9. Experimental results of HAT and CHAT at scale ×4 on partial X-Ray dataset.**

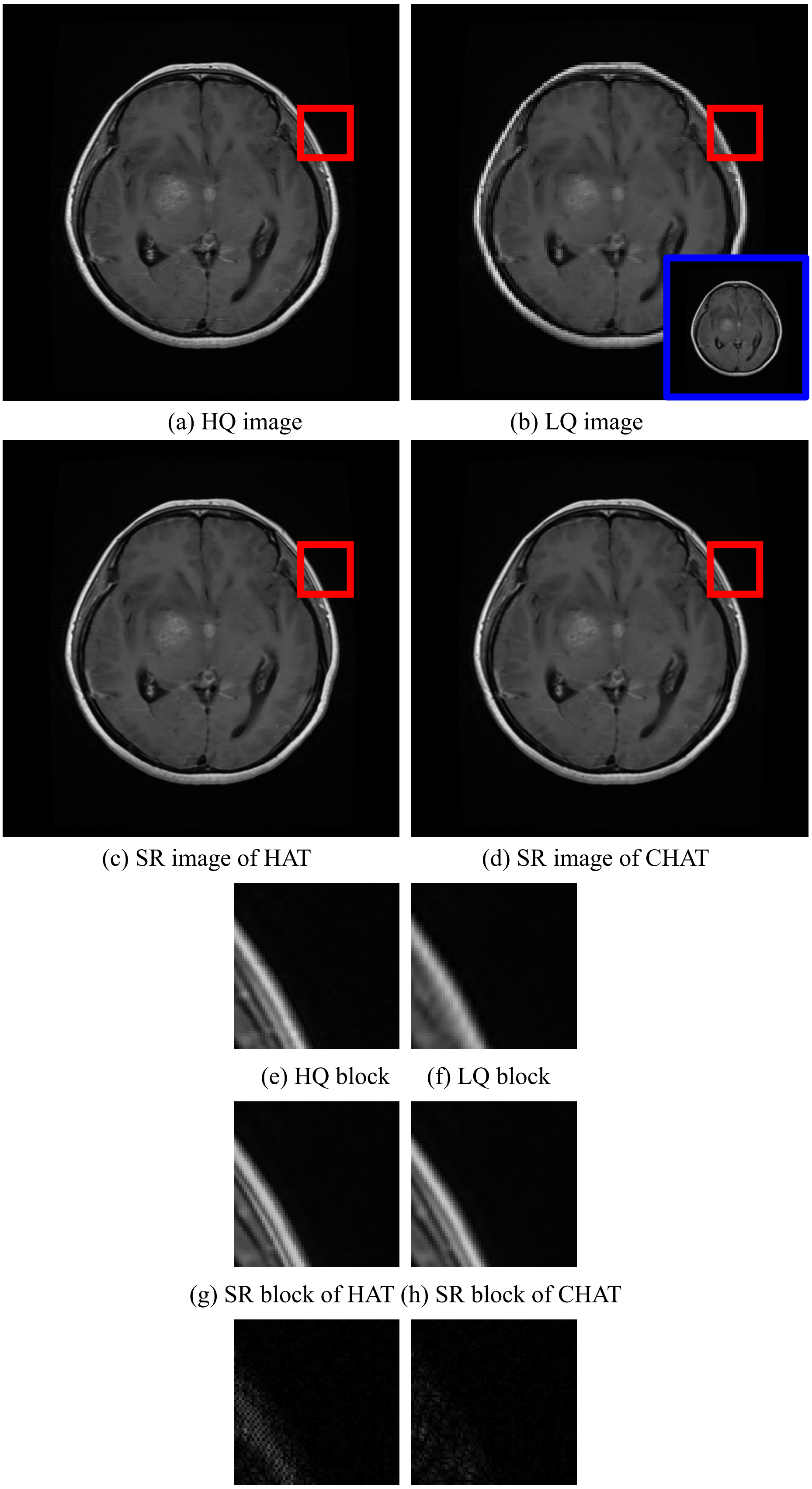

(a) HQ image (b) LQ image

(c) SR image of HAT (d) SR image of CHAT

(e) HQ block (f) LQ block

(g) SR block of HAT (h) SR block of CHAT

(i) difference block of HAT (j) difference block of CHAT

**Figure 10. Experimental results of HAT and CHAT at scale ×3 on partial MRI dataset.**

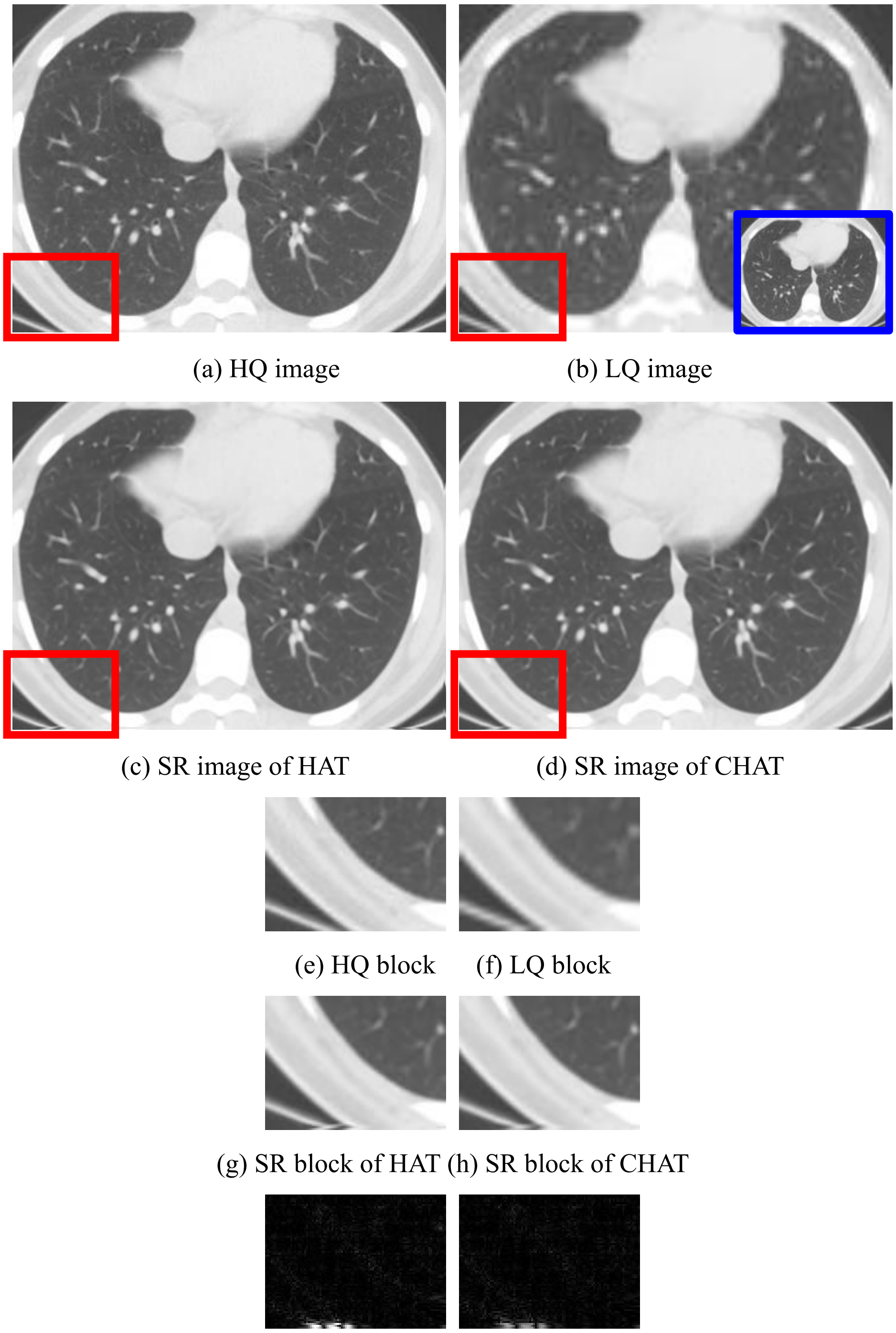

(a) HQ image (b) LQ image

(c) SR image of HAT (d) SR image of CHAT

(e) HQ block (f) LQ block

(g) SR block of HAT (h) SR block of CHAT

(i) difference block of HAT (j) difference block of CHAT

**Figure 11. Experimental results of HAT and CHAT at scale ×3 on partial CT dataset.**

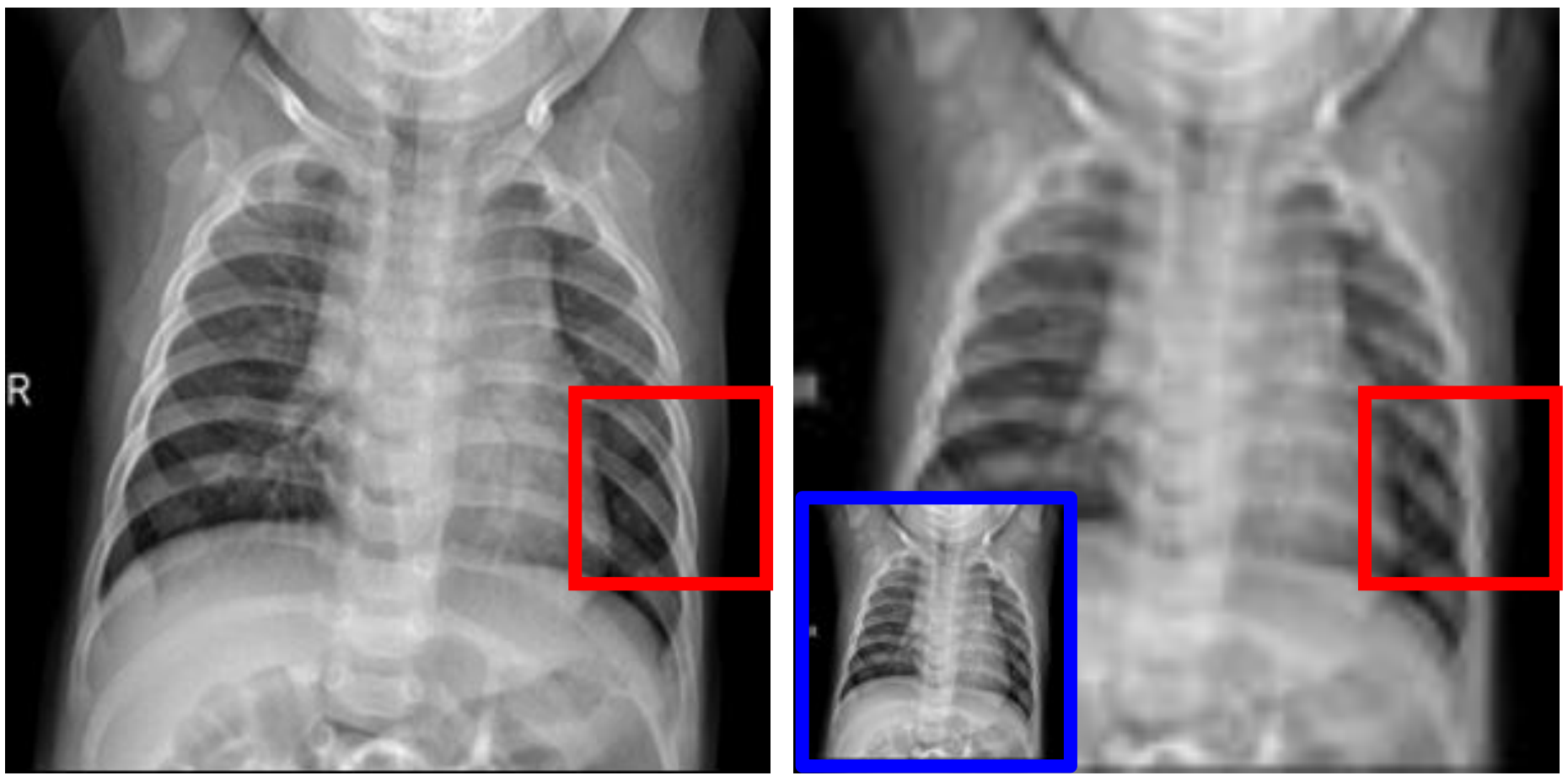

(a) HQ image (b) LQ image

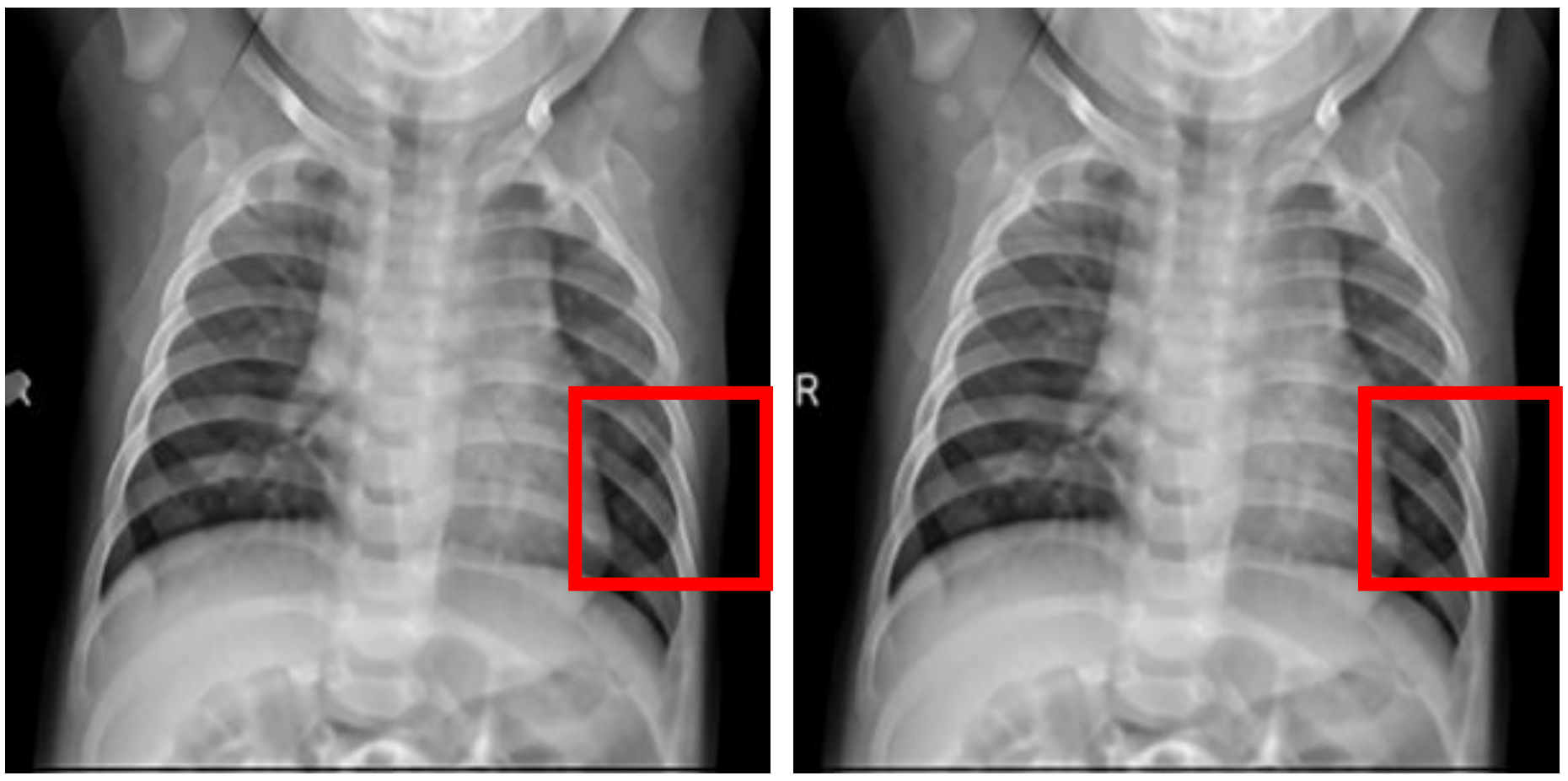

(c) SR image of HAT (d) SR image of CHAT

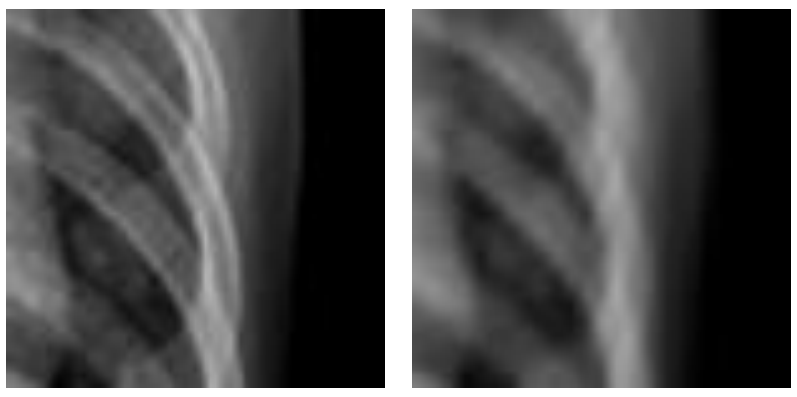

(e) HQ block (f) LQ block

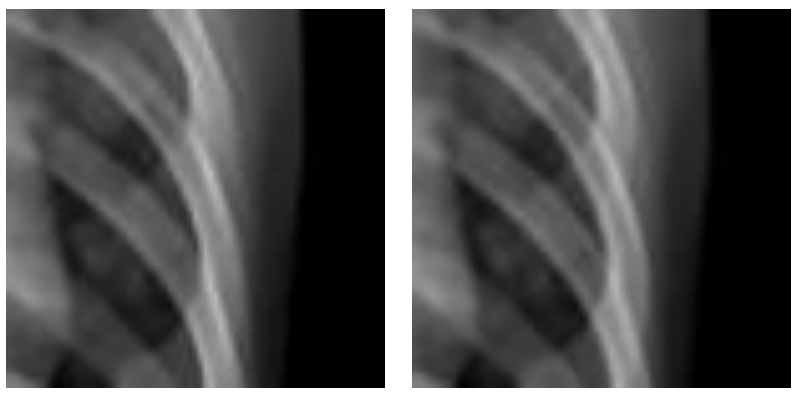

(g) SR block of HAT (h) SR block of CHAT

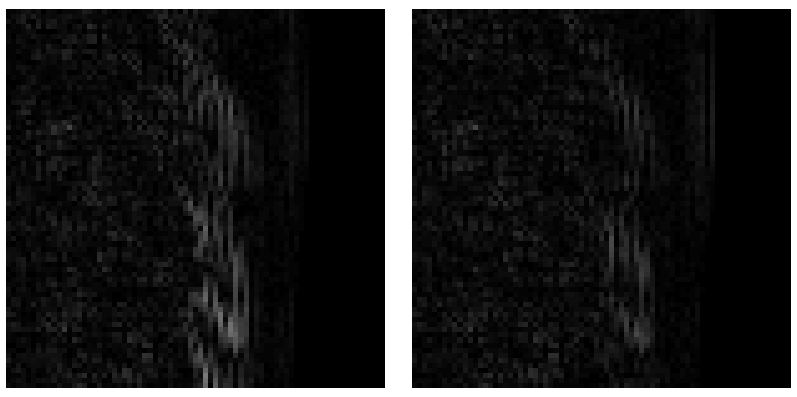

(i) difference block of HAT (j) difference block of CHAT

**Figure 12. Experimental results of HAT and CHAT at scale ×3 on partial X-Ray dataset.**

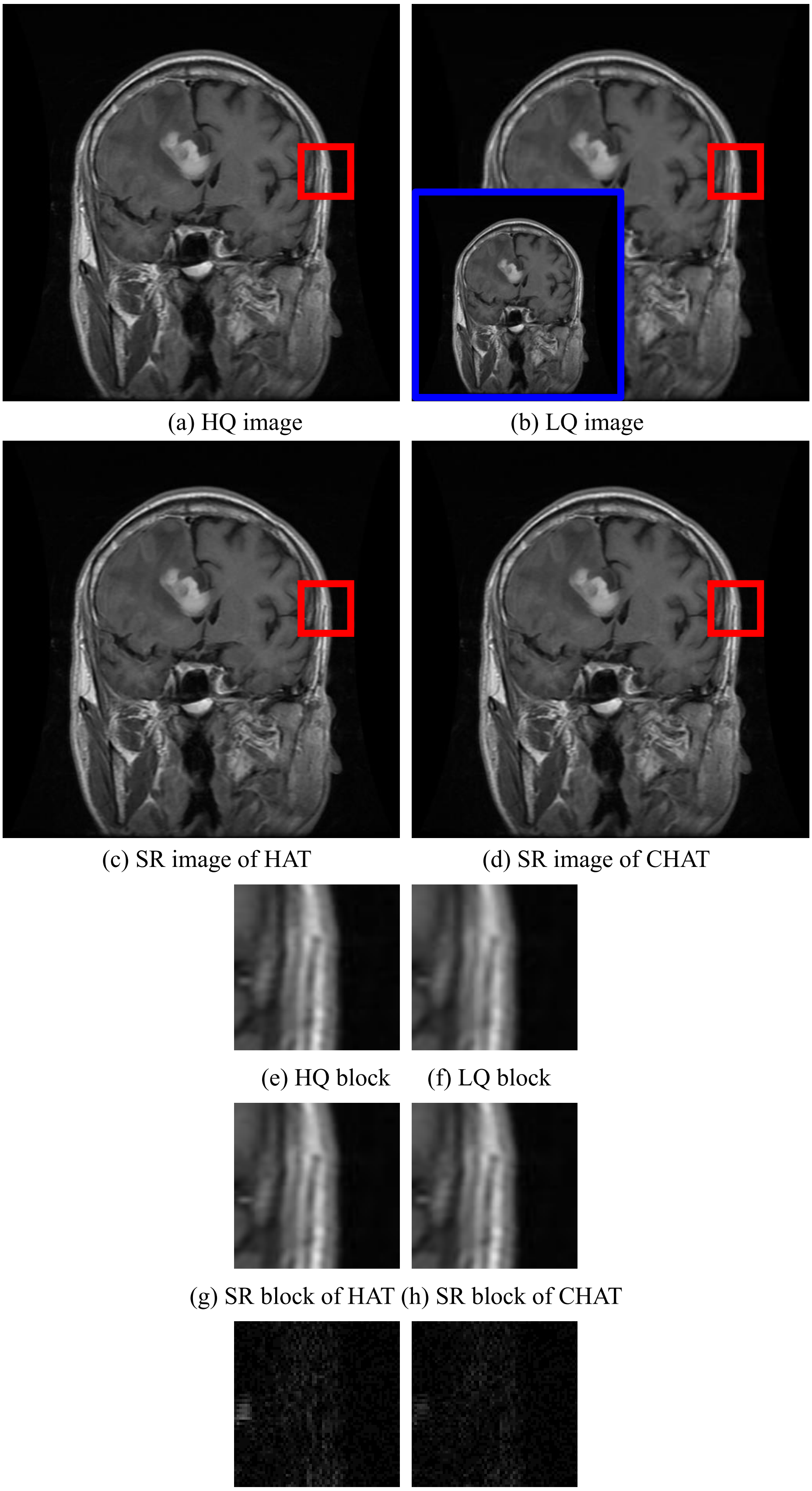

(a) HQ image (b) LQ image

(c) SR image of HAT (d) SR image of CHAT

(e) HQ block (f) LQ block

(g) SR block of HAT (h) SR block of CHAT

(i) difference block of HAT (j) difference block of CHAT

**Figure 13. Experimental results of HAT and CHAT at scale ×2 on partial MRI dataset.**

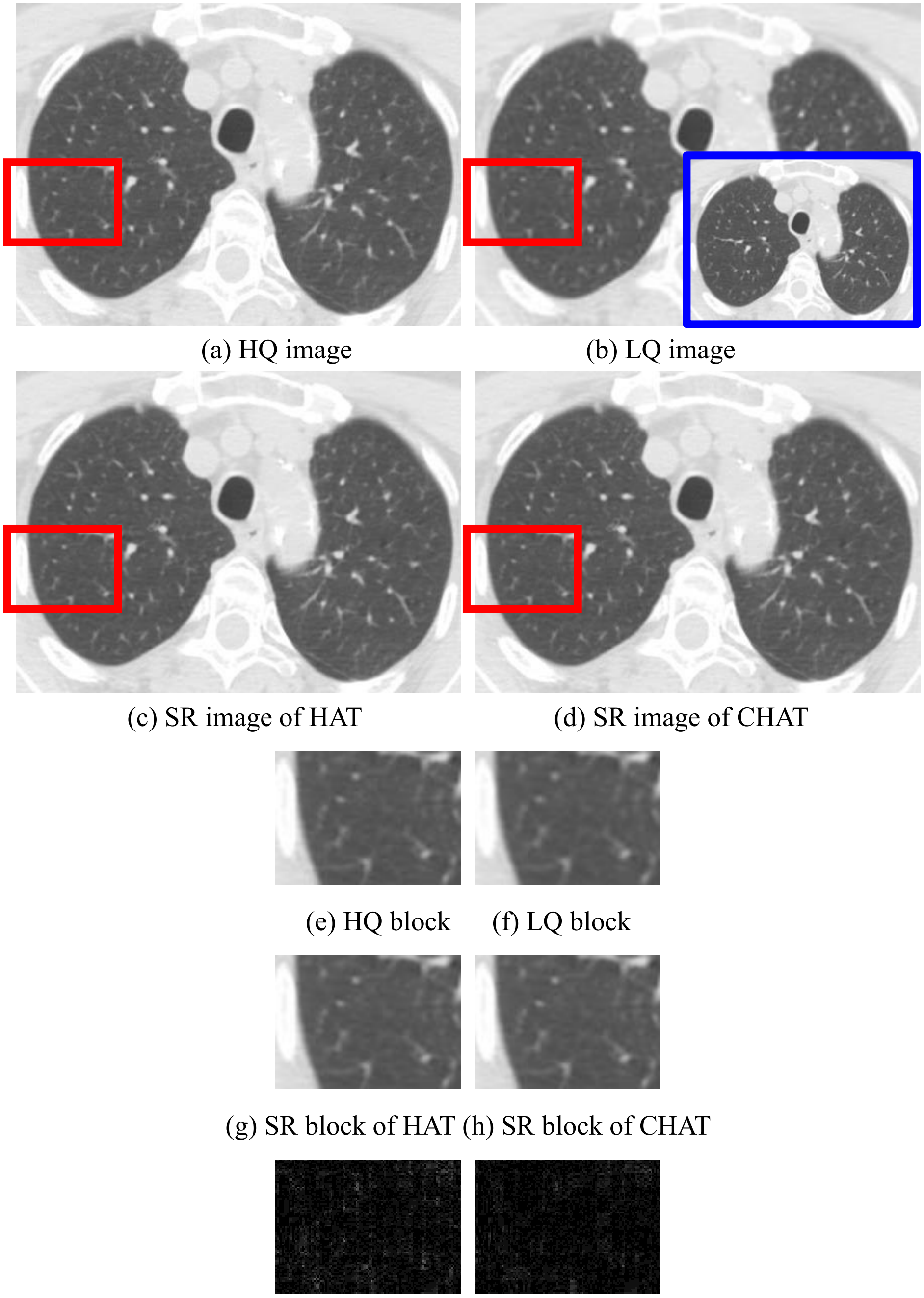

(a) HQ image (b) LQ image

(c) SR image of HAT (d) SR image of CHAT

(e) HQ block (f) LQ block

(g) SR block of HAT (h) SR block of CHAT

(i) difference block of HAT (j) difference block of CHAT

**Figure 14. Experimental results of HAT and CHAT at scale ×2 on partial CT dataset.**

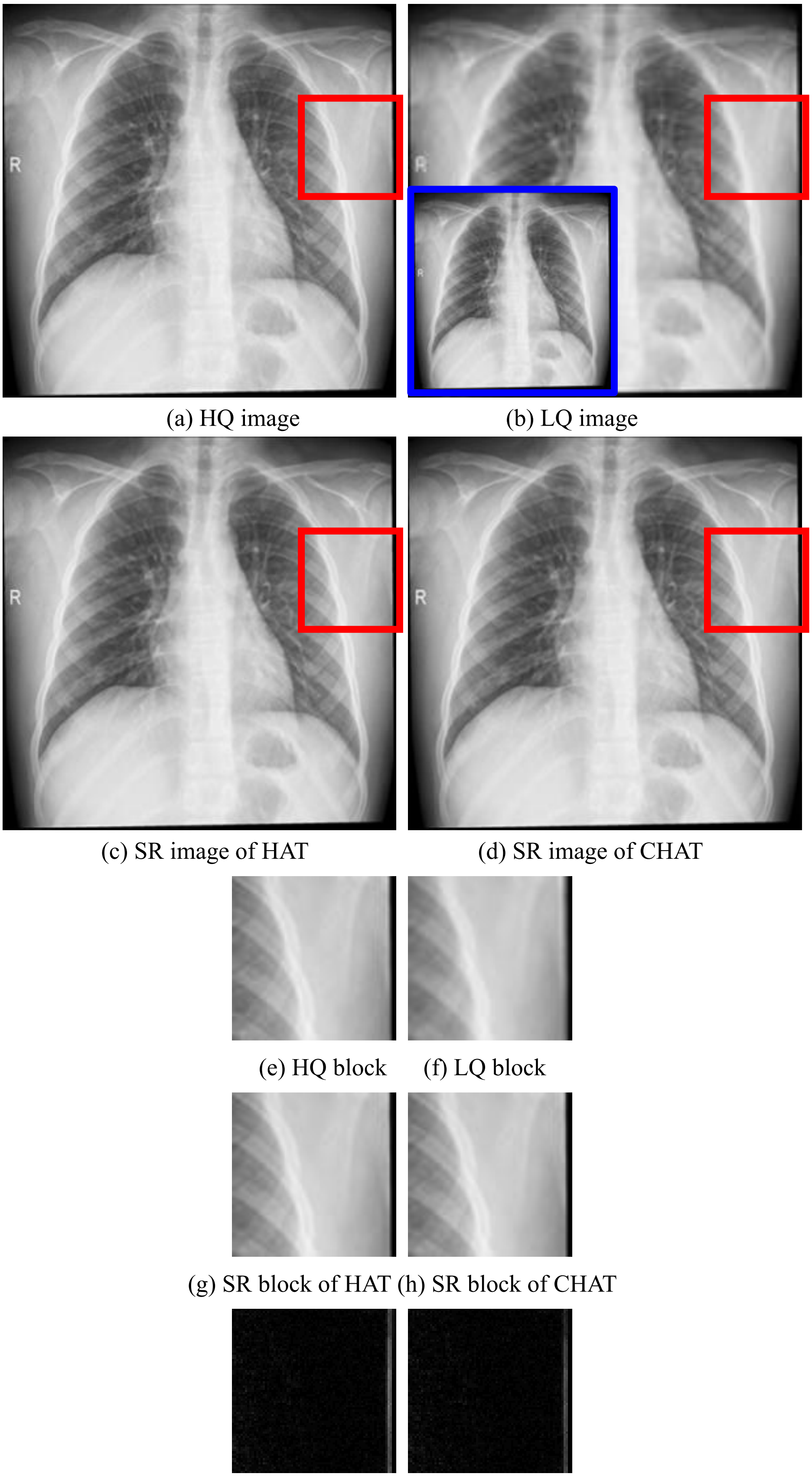

(a) HQ image (b) LQ image

(c) SR image of HAT (d) SR image of CHAT

(e) HQ block (f) LQ block

(g) SR block of HAT (h) SR block of CHAT

(i) difference block of HAT (j) difference block of CHAT

**Figure 15. Experimental results of HAT and CHAT at scale ×2 on partial X-Ray dataset.**

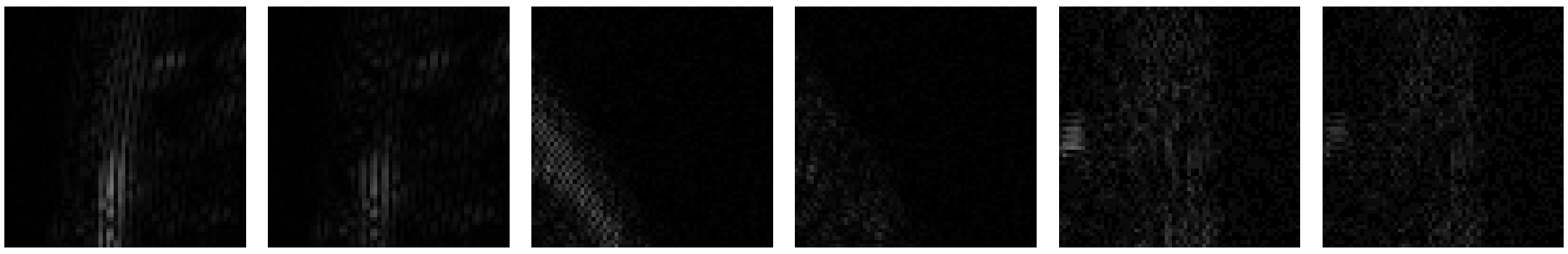

HAT,×4 CHAT,×4 HAT,×3 CHAT,×3 HAT,×2 CHAT,×2

(a) partial MRI dataset

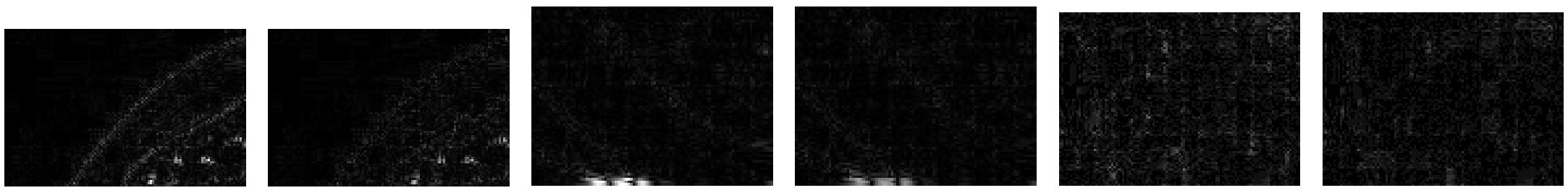

HAT,×4 CHAT,×4 HAT,×3 CHAT,×3 HAT,×2 CHAT,×2

(b) partial CT dataset

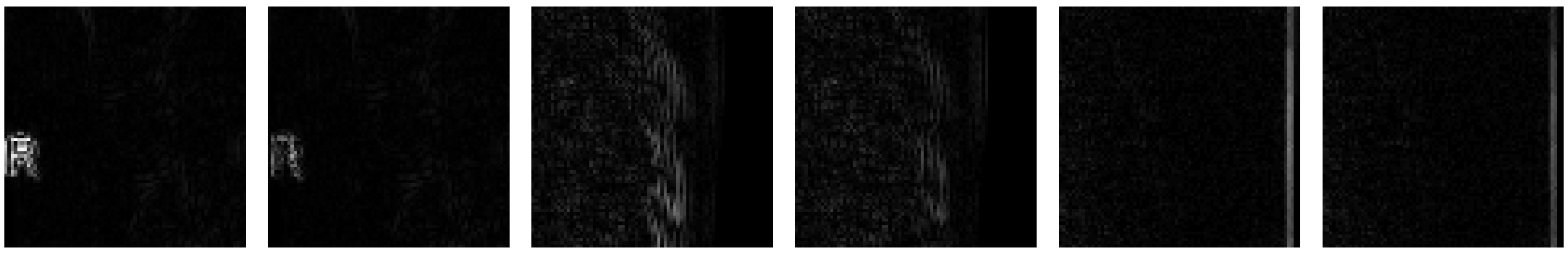

HAT,×4 CHAT,×4 HAT,×3 CHAT,×3 HAT,×2 CHAT,×2

(c) partial X-RAY dataset

**Figure 16. Difference blocks of HAT and CHAT at three scales on three partial datasets.**

For the sake of verifying the performance of the proposed methods on large datasets, the experimental results of two representative algorithms, HAT and CHAT, on three full datasets are enumerated in Table 10. It is displayed in Table 10 that CHAT distinctly outperforms HAT at three image scales (x2, x3, and x4), on three full datasets, and in PSNR, SSIM, and FID.

For the purpose of comparing the performance of different algorithms, PSNR, SSIM, and FID at three scales on the partial MRI dataset from Tables 5 to 9 are respectively shown in Figures 4 to 6. It is disclosed in Figure 4 that the proposed CMISR outperforms classical open-loop ISR in reconstruction performance. It is also disclosed in Figure 4 that CHAT has the best PSNR at scales x3 and x4, and CDAT has the best PSNR at scale x2. It is reflected in Figure 5 that the proposed CMISR demonstrates superiority over classical open-loop ISR in reconstruction capability. It is also reflected in Figure 5 that CHAT holds the best SSIM at scales x3 and x4, and CDAT holds the best SSIM at scale x2. It is indicated in Figure 6 that the proposed CMISR outperforms classical open-loop ISR in reconstruction capacity. It is also indicated in Figure 6 that CHAT possesses the best FID at scale x4, CSRFormer possesses the best FID at scale x3, and CDAT possesses the best FID at scale x2.

In order to visually compare the performance of HAT and CHAT, Figures 7 to 15 demonstrate some example images at three scales and on three partial datasets. Figures 7 to 9 are the exemplar images at scale x4 on three partial datasets, Figures 10 to 12 are the exemplar images at scale x3 on three partial datasets, and Figures 13 to 15 are the exemplar images at scale x2 on three partial datasets.

In Figure 7, subfigure (a) is the original HQ image, subfigure (b) is the LQ image, subfigure (c) is the SR reconstruction image of HAT, and subfigure (d) is the SR reconstruction image of CHAT. For the convenience of comparison, subfigure (b) is enlarged to the same size as subfigure (a). The subfigure with a blue box on subfigure (b) is the LQ image without enlargement.

It is uneasy for human visual systems to discriminate subfigures (a) to (d). Hence, small blocks in subfigures (a) to (d) are chosen for clear comparison. The small blocks in subfigures (a) to (d) are marked with red boxes. Subfigures (e) to (h) are the zoomed versions of the small blocks.

It is still uneasy for human visual systems to discriminate subfigures (e) to (h). Thus, a difference block is computed for further clear comparison. The difference block is the absolute difference between the SR reconstruction block and the HQ block. Subfigure (i) is the zoomed version of the difference block between subfigures (g) and (e). Subfigure (j) is the zoomed version of the difference block between subfigures (h) and (e). The obvious difference can be found between subfigures (i) and (j).

Analogously to Figure 7, Figures 8 to 15 show the HQ images, the LQ images, the SR images, and corresponding difference blocks at different scales on different partial datasets. It is indicated in Figures 7 to 15 that the proposed CHAT prevails over HAT in the capability of SR reconstruction.

It should be mentioned that there is a character 'R' in Figure 9. It is displayed in Figure 9 that CHAT possesses a higher capability of SR reconstruction for the character in the medical image than HAT. This situation can also be found in Figure 15 which has black borders. The character and block border have strong edges, so CHAT is more suitable for medical images with strong edges or intense contrast than HAT.

Similarly to the experimental results of CHAT and HAT in Figures 7 to 15, it can be easily deduced that the proposed CSRFormer, CDAT, CEMT, and CSwinIR outperform classical SRFormer, DAT, EMT, and SwinIR in the performance of SR reconstruction.

In order to clearly compare the performance of HAT and CHAT at three scales on three partial datasets, the difference blocks in Figures 7 to 15 are merged into Figure 16.

All in all, the proposed closed-loop CMSIR is superior to classical open-loop ISR in the capacity of SR reconstruction.

# 5 Conclusion

This paper presents a closed-loop CMISR with explicit UR and SR units. The UR unit can always be given, assumed, or estimated. The proposed CMISR adopts a single, global, and intra-image-based FB mechanism. The mathematical model of CMISR is established, the closed-loop equation of CMISR is achieved, and the mathematical analysis with Taylor-series approximation is conducted. The proposed CIMSR holds zero recovery error in steady-state. The proposed CMISR also holds the property of plug-and-play which merges model-based and learning-based approaches and can be built on any existing advanced ISR algorithms. Based on five state-of-the-art ISR algorithms, five CMISR algorithms are respectively raised: CHAT, CSRFormer, CDAT, CEMT, and CSwinIR. Simulation experiments are implemented with three scale factors and on three open medical image datasets, MRI, CT, and X-ray datasets. Experimental results show the proposed CMISR surpasses classical open-loop ISR in the SR reconstruction performance, such as PSNR, SSIM, and FID. In addition, the proposed CMISR is particularly well-suited to medical images with sharp edges or high contrast.

In our future work, the UR element of blind and real-world CMISR will be explored. The degradation model of UR can be built with or without degradation kernel estimation via learning or model-based methods. The closed-loop architecture can be incorporated into the loss functions of deep learning. The FB mechanism in the training phase, including network architecture and loss function, of deep learning will also be investigated, and the experiments on more medical image datasets will further be undertaken.

# Author Contributions

Conceptualization, Honggui Li and Maria Trocan; Methodology, Honggui Li and Dimitri Galayko; Writing, Honggui Li and Nahid Md Lokman Hossian; Supervision, Mohamad Sawan. All authors have read and agreed to the published version of the manuscript.

# Competing Interests

The authors declare that they have no known competing financial interests or personal relationships that could have appeared to influence the work reported in this paper.

# Acknowledgment

The authors would very much like to thank all the authors of the 5 competing algorithms for selflessly releasing their source codes of image super-resolution on the Microsoft GitHub website. The open-source codes allow us to easily implement the proposed algorithm depending on the competing algorithms. The authors also would very much like to express deep thanks to Google COLAB for its free GPU computing service.

## Biographies

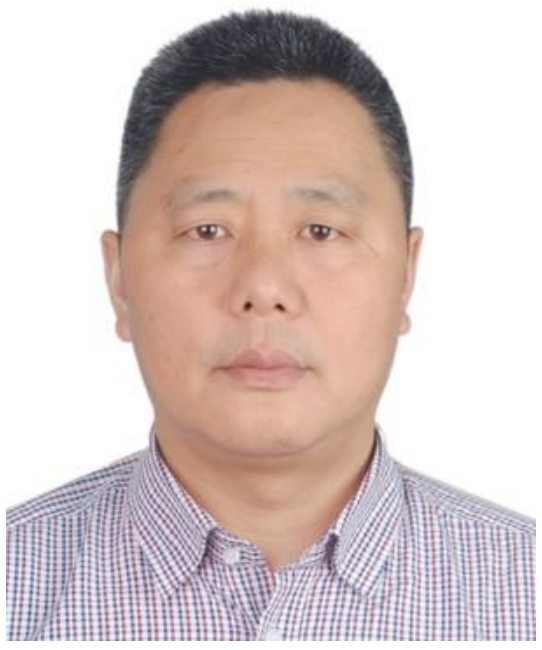

**Honggui Li** received a B.S. degree in electronic science and technology from Yangzhou University and received a Ph.D. degree in mechatronic engineering from Nanjing University of Science and Technology. He is a senior member of the Chinese Institute of Electronics. He is a visiting scholar and a post-doctoral fellow at Institut Supérieur d'Électronique de Paris for one year. He is an associate professor of electronic science and technology and a postgraduate supervisor of electronic science and technology at Yangzhou University. He is a reviewer for some international journals and conferences. He is the author of over 30 refereed journal and conference articles. His current research interests include machine learning, deep learning, computer vision, and embedded computing.

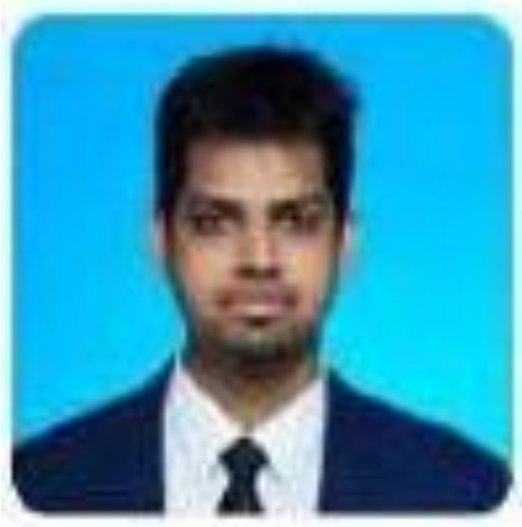

**Nahid Md Lokman Hossain** received a B.S. degree in computer science and technology from Chongqing University of Posts and Telecommunications in China. He is now studying for a master degree in software engineering at Yangzhou University in China. His research interests include machine learning, computer vision, innovation software, cloud computing, and big data.

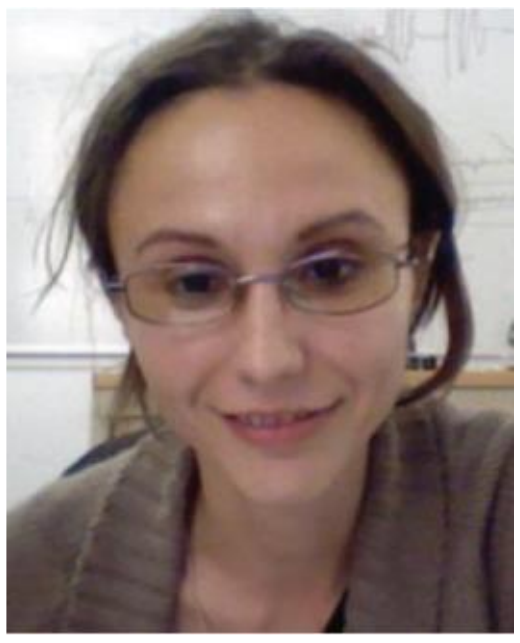

**Maria Trocan** received a M.Eng. in Electrical Engineering and Computer Science from the Politehnica University of Bucharest, a Ph.D. in Signal and Image Processing from Telecom ParisTech, and the Habilitation to Lead Researches (HDR) from Pierre & Marie Curie University (Paris 6). She has joined Joost - Netherlands, where she worked as a research engineer involved in the design and development of video transcoding systems. She is firstly Associate Professor, then Professor at Institut Superieur d'Electronique de Paris (ISEP). She is an Associate Editor for the Springer Journal on Signal, Image and Video Processing and a Guest Editor for several journals (Analog Integrated Circuits and Signal Processing, IEEE Communications Magazine, etc.). She is an active member of IEEE France and served as a counselor for the ISEP IEEE Student Branch, IEEE France Vice-President responsible for Student Activities, and IEEE Circuits and Systems Board of Governors member, as Young Professionals representative. Her current research interests focus on image and video analysis & compression, sparse signal representations, machine learning, and fuzzy inference.

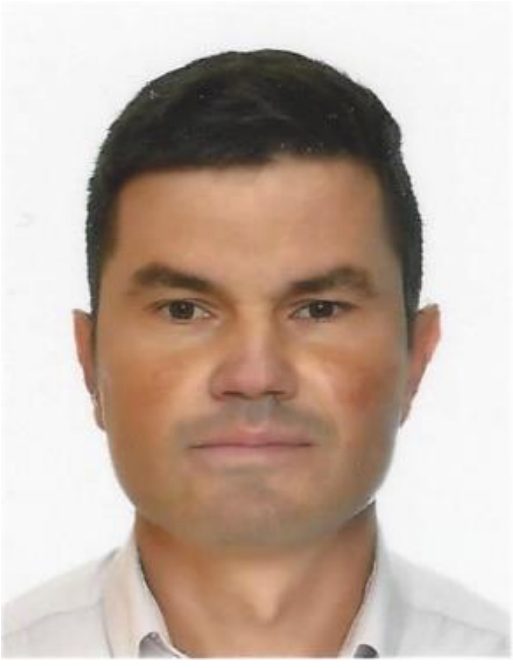

**Dimitri Galayko** received a bachelor's degree from Odessa State Polytechnic University in Ukraine, a master's degree from the Institute of Applied Sciences of Lyon in France, and a Ph.D. degree from University Lille in France. He made his Ph.D. thesis at the Institute of Microelectronics and Nanotechnologies. His Ph.D. dissertation was on the design of micro-electromechanical silicon filters and resonators for radio-communications. He is a Professor at the LIP6 research laboratory of Sorbonne University in France. His research interests include the study, modeling, and design of nonlinear integrated circuits for sensor interfaces and mixed-signal applications. His research interests also include machine learning and fuzzy computing.

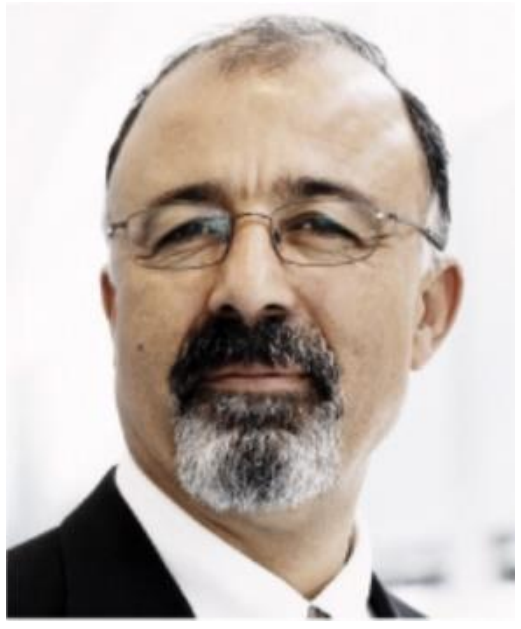

**Mohamad Sawan** (Fellow, IEEE) received a Ph.D. degree in electrical engineering from the University of Sherbrooke, Sherbrooke, QC, Canada, in 1990. He was a Chair Professor awarded with the Canada Research Chair in Smart Medical Devices (2001–2015) and was leading the Microsystems Strategic Alliance of Quebec - ReSMiQ (1999–2018). He is a Professor of Microelectronics and Biomedical Engineering, in leave of absence from Polytechnique Montréal, Canada. He joined Westlake University, Hangzhou, China, in January 2019, where he is a Chair Professor, Founder, and the Director of the Center for Biomedical Research And Innovation (CenBRAIN). He has published more than 800 peer-reviewed articles, two books, ten book chapters, and 12 patents. He founded and chaired the IEEE-Solid State Circuits Society Montreal Chapter (1999–2018) and founded the Polystim Neurotech Laboratory, Polytechnique Montréal (1994–present), including two major research infrastructures intended to build advanced Medical devices. He is the Founder of the International IEEE-NEWCAS Conference, and the Co-Founder of the International IEEE-BioCAS, ICECS, and LSC conferences. He is a Fellow of the Royal Society of Canada, a Fellow of the Canadian Academy of Engineering, and a Fellow of the Engineering

Institutes of Canada. He is also the "Officer" of the National Order of Quebec. He has served as a member of the Board of Governors (2014–2018). He is the Vice-President of Publications (2019–present) of the IEEE CAS Society. He received several awards, among them the Queen Elizabeth II Golden Jubilee Medal, the Barbara Turnbull 2003 Award for spinal-cord research, the Bombardier and Jacques-Rousseau Awards for academic achievements, the Shanghai International Collaboration Award, and the medal of merit from the President of Lebanon for his outstanding contributions. He was the Deputy Editor-in-Chief of the IEEE TRANSACTIONS ON CIRCUITS AND SYSTEMS-II: EXPRESS BRIEFS (2010–2013); the Co-Founder, an Associate Editor, and the Editor-in-Chief of the IEEE TRANSACTIONS ON BIOMEDICAL CIRCUITS AND SYSTEMS; an Associate Editor of the IEEE TRANSACTIONS ON BIOMEDICALS ENGINEERING; and the International Journal of Circuit Theory and Applications.